%% file: paper.tex
\documentclass[manuscript,screen,nonacm]{acmart}

\AtBeginDocument{%
  \providecommand\BibTeX{{%
    \normalfont B\kern-0.5em{\scshape i\kern-0.25em b}\kern-0.8em\TeX}}}

\usepackage{graphicx}
\usepackage{algorithmic}
\usepackage{amsmath,amsfonts} 
\usepackage[english]{babel}
\usepackage{todonotes}
\usepackage{lscape}
\usepackage{multirow}
\usepackage{paralist}
\usepackage[figuresright]{rotating} 
\usepackage{tabularx} 
\usepackage{textcomp} 
\usepackage{xcolor} 
\usepackage{enumitem}

\usepackage{subfig} 

\input{./pictures/system_types.tex}

\begin{document}

\title{Towards Polyglot Data Stores}
\subtitle{Overview and Open Research Questions}

\author{Daniel Glake}
\authornote{These authors contributed equally to this research.}
\email{daniel.glake@uni-hamburg.de}
\author{Felix Kiehn}
\authornotemark[1]
\email{felix.kiehn@uni-hamburg.de}
\author{Mareike Schmidt}
\authornotemark[1]
\email{mareike.schmidt-3@uni-hamburg.de}

\affiliation{%
  \institution{Universität Hamburg}
}

\author{Fabian Panse}
\email{fabian.panse@uni-hamburg.de}
\author{Norbert Ritter}
\email{norbert.ritter@uni-hamburg.de}

\affiliation{%
  \institution{Universität Hamburg}
}

\authorsaddresses{%
Authors’ address: Daniel Glake, daniel.glake@uni-hamburg.de; Felix Kiehn, felix.kiehn@uni-hamburg.de; Mareike Schmidt, mareike.schmidt-3@uni-hamburg.de; Fabian Panse, fabian.panse@uni-hamburg.de; Norbert Ritter, norbert.ritter@uni-hamburg.de, Department of Informatics, Universität Hamburg, Vogt-Kölln-Straße 30, 22846 Hamburg, Germany.
}

\renewcommand{\shortauthors}{Glake, Kiehn, Schmidt, Panse, and Ritter}

\begin{abstract}
Nowadays, data-intensive applications face the problem of handling heterogeneous data with sometimes mutually exclusive use cases and soft non-functional goals such as consistency and availability. Since no single platform copes everything, various stores (RDBMS, NewSQL, NoSQL) for different workloads and use-cases have been developed. However, since each store is only a specialization, this motivates progress in polyglot data management emerged new systems called Mult- and Polystores. They are trying to access different stores transparently and combine their capabilities to achieve one or multiple given use-cases.
This paper describes representative real-world use cases for data-intensive applications (OLTP and OLAP). It derives a set of requirements for polyglot data stores. Subsequently, we discuss the properties of selected Multi- and Polystores and evaluate them based on given needs illustrated by three common application use cases. We classify them into functional features, query processing technique, architecture and adaptivity and reveal a lack of capabilities, especially in changing conditions tightly integration. Finally, we outline the benefits and drawbacks of the surveyed systems and propose future research directions and current challenges in this area.
\end{abstract}



\maketitle

\section{Introduction}
After decades of dominance of relational database management systems (DBMS) since the 70's, the appearance of a multitude of new data stores, coined NoSQL (Not only SQL) systems, since 2009 changed the database and information systems landscape immensely.
The emergence of ever-growing amounts of (often unstructured) data - particularly in the world of web applications - overcharged the technical capabilities of monolithic relational DBMS \cite{gessert2017nosqltoolbox, mazumder2016, curino2011relational} that tried to handle the expanding workloads by using highly specialized hardware like Field Programmable Gate Array and Graphics Processing Units \cite{bress2014gpu, bakkum2010accelerating}. 

Even though the resulting landscape of diverse stores enables users to select an appropriate system based on their specific requirements, it poses several problems \cite{sadalage2013nosql}:
\begin{itemize}[topsep=-0.5pt,itemsep=-2pt] 
	\item Even as an expert, it is hard to keep an overview of all these stores and their respective sweet spots and limitations (even if some guidelines \cite{gessert2017nosqltoolbox} are available).
	\item Complex applications may have subcomponents with contradictory requirements so that using a single data persistence solution always ends up in a trade-off between them.
	\item As applications change and evolve, so do their requirements. These changes are often not immediately
	visible to the user so that the initially used data store may not remain the best fitting one over time.
\end{itemize}
The approach of \emph{polyglot persistence} \cite{sadalage2013nosql} aims to solve these problems by combining the benefits of several data stores (and their underlying techniques) without adopting their drawbacks and to hide the automatic coordination across these (heterogenous-)stores behind single or multiple interfaces.

Of course, polyglot persistence comes with a lot of challenges
because the system does not only has to orchestrate different stores based on dynamically changing conditions,
but also has to deal with mismatches between different data models, 
and needs to translate and mediate queries.

The objective of this paper is to survey the first generation of polyglot data stores. This includes descriptions and discussions on several aspects such as architecture and query processing approach.

The main contributions of this paper can be summarized as follows:
\begin{itemize} [topsep=-0.5pt,itemsep=-2pt] 
    \item A description of three real-world use cases that motivate the use of polyglot data management.
	\item A list of functional and non-functional features that may be provided by a polyglot data store in order to solve representative real world use cases.
	\item A description and detailed comparison of currently existing polyglot data stores.
	\item An overview of open challenges in polyglot data management.
\end{itemize}

The remainder of this paper is structured as follows: In Section~\ref{sec:Motivation}, we motivate the use of polyglot data management by different use cases, illustrating different requirements for data management. Given these requirements, we describe and discuss technical features in Section~\ref{sec:Concluded Requirements}, leading to a range of polyglot data management kinds in Section~\ref{sec:System types}. In Section~\ref{sec:Analyzed Systems} we survey existing systems that can be assigned to any of these system kinds and describes established polystores like BigDAWG~\cite{duggan2015bigdawg} or upcoming systems like Polypheny-DB~\cite{vogt2018polypheny,vogt2020polypheny}. In detail, we describe each system's specialization and approach to plan, optimize, and execute queries and what kind of systems are utilized in this process. Section~\ref{sec:Comparison} give the complete overview of all considered systems and reveals features commonly used by most of the systems and aspects they are still missing. These missing features or problems are then discussed in Section~\ref{sec:Open Challenges} before Section~\ref{sec:Conclusion} concludes the paper.

\section{Motivation}\label{sec:Motivation}

In this section, we address three different use-cases from which we illustrated relevant requirements in the industry features required in concrete data management design. We consider one use case focusing more on a transactional workload and processing (OLTP), whereas the other examines challenges in processing analytical workloads (OLAP) and another in dealing a hybrid transactional and analytical processing.

\subsection{E-Commerce and Customer Management}\label{sec:UseCaseECommerce}
Consider a well-known online store that targets international markets \cite{zhang2014efficient, hasselbring2017microservice, gessert2015polyglot}. Such a system promotes ordinary consumer goods using text descriptions, ratings, short video and image data with  non-functional requirement in order to serve purchasing services all the time. From a business perspective, user groups may be interested in changing sales volumes, prices per article or other key indicators.
This context represents a well suited transactional processing (OLTP) use case, in which a set of system specifications have to be handled:

\begin{compactenum}[E1.]
	\item \label{item: ECom_Availability} \emph{Availability}: High availability for sales operations is a crucial requirement for e-commerce systems as downtimes lead to a loss in earnings~\cite{anthony2006market} and come to a trade-off between consistency and availability. The decision can be made to occasionally accept downtimes such as customer support or article rating instead of shut-down purchasing features.
	\item \label{item: ECom_FlexibleConsistency} \emph{Flexible Consistency}: On a state-of-the-art, modular e-commerce website, each component requires a different level of consistency (e.g. high consistency for payments, low consistency for product ratings).
	\item \label{item: ECom_ReadThroughput} \emph{Read Throughput}: When navigating through the website, users request more and more articles and other data in the system. Therefore, a read-friendly distribution model (e.g. replicating a subset of data node) or a reschedule for better-performing hardware needs to be supported.
	\item \label{item: ECom_ComplexDataAnalysis} \emph{Complex Analyses}: 
	To enable analytical tasks such as calculating user-specific product recommendations
	or analyzing the company's recent business numbers,
	data mining and OLAP querying capabilities have to be provided.
\end{compactenum}

\subsection{Agent-Based Simulation and Decision Support System}\label{sec:UseCaseDss}

Another real-world use-case considers the widely used form of a mobility simulation and digitalized cities called digital twins \cite{clemen2021multi, glake2021data, glake2020utilizing} such as representations of Hamburg \cite{weyl2018agent} ecology systems in Africa \cite{lenfers2021improving}.
Citizens have attributes (e.g. \emph{velocity} and \emph{position}) and move along a road network affected by traffic lights, intersections and other participants. Depending on the scenario's selected time-scale and spatial extension, the simulation produces an extensive set of varying simulation results. The output includes hierarchically structured spatial-temporal data and interrelationships between agents. Due to the multiplicity of results, formats and analytical routines, this simulation is a prime example for analytical processing (OLAP), where the following system properties are required:
\begin{compactenum}[S1.]
	\item \label{item: Sim_WriteThroughput} \emph{Write Throughput}: 
	When executing a mobility simulation incorporating many agents such as modelled as citizens on the street, many agent and environmental states are computed in each simulation step. A scaled solution is required to analyze all these states and visualize, for example, all computed trajectories in the world. These solutions have to handle massive write processes in terms of data loaded into the simulation and computed by the simulation.
	\item \label{item: Sim_ImmuntableData} \emph{Immutable Data}: Scientific users as the primary target group are convinced that data should never be discarded. Even if data objects are incorrect, their revision should never overwrite old data since previously published works may have used them. 
	Instead, we need a versioning approach that enables users to selectively access individual versions.
	\item \label{item: Sim_SupportSpatialFormats} \emph{Support of Spatial Formats}: For a geographic-driven simulation, it is mandatory to define standard spatial formats and allow corresponding queries (e.g. spatial-joins, k-nearest neighbours).
	\item \label{item: Sim_RealTimeAnalysis} \emph{Real-time Analyses}: In order to display aggregated indicators as well as the movement of agents on a map, data objects are streamed through the system at runtime, when data has been changed (push-based queries).
	\item \label{item: Sim_ComplexAnalysis} \emph{Complex Analysis}: In addition to simple queries and data browsing along the different dimensions \cite{glake2021hierarchical}, various kinds of complex analyses such as k-means and correlation analysis are beneficial in a scientific context. For the case of traffic simulations where graphs are used to represent the road network, graph analyzes should be possible as well.
\end{compactenum}

\subsection{Healthcare Data Management Systems}\label{sec:UseCaseMed}
An often described use case for polyglot data management systems is data in healthcare as provided by the MIMIC II ("Multiparameter Intelligent Monitoring in Intensive Care II") database \cite{saeed2011multiparameter, elmore2015demonstration, kaur2015smart}. It contains detailed information on patients, laboratory and radiology results, physiological data (e.g. electrocardiograms) as well as doctor's and nurse's notes relating observations and treatments. All these different kinds of information have to be gathered and analysed in common modern hospitals.

\begin{compactenum}[M1.]
    \item \label{item: Healthcare_MultipleDataFormats} \emph{Support of Multiple Data Formats:} Data in this medical use case ranges from structured and unstructured data (patient data, lab results) over images and waveform data (CT and MRI images, electrocardiograms) to graph based data (relationships between patients or blood relations).
    \item \label{item: Healthcare_ComplexAnalysis} \emph{Complex Analyses:} The analysis of images and waveform data in the case of medical imaging, requires complex operations such es Fourier- and Wavelet-Transformations as well as further image analysis capabilities (e.g. to find regions containing abnormal tissues). Furthermore, text search and graph algorithms may be required (e.g. in order to find similar examination reports or to analyse relationships between patients).
    \item \label{item: Healthcare_RealTime} \emph{Real-time Support:} In order to keep patients in an intensive care unit under continuous  surveillance and detect, for example, abnormal heart rhythms in time, real-time monitoring is essential.
    \item \label{item: Healthcare_Availability} \emph{Availability:} In case of emergencies, it is necessary that requested data such as the patient's allergies or blood type are available at any time.
    \item \label{item: Healthcare_Consistency} \emph{Consistency:} The treatment of patients always have to follow the latest diagnosis. Accessing outdated data might even endanger the patients' lives. Therefore, a high level of consistency is needed for specific parts of the medical data.
    \item \label{item: Healthcare_Security_Privacy} \emph{Security and Privacy:} In a healthcare data management system, a huge amount of personal data is collected and it is highly important, that the data is protected from illegal access. Additionally, it should be possible to render data anonymous in order to use it for research purposes.
\end{compactenum}

\section{Concluded Requirements and System Characteristics}\label{sec:Concluded Requirements}

From the use cases in Section~\ref{sec:Motivation}, we can derive a variety of requirements data management systems have to fulfil in order to provide a sufficient persistence solution. In this section, we assemble a list of these concluded requirements and discuss them in more detail. Here we distinguish between requirements that are directly concluded from the use cases and requirements that are induced from the direct ones.

\subsection{Explicit Requirements}\label{subsec:Functional Features}

From our representative use cases, we derive seven essential, indispensable \emph{functional} features, which needs to be support by a future-proof polyglot data store.

\textbf{Modeling Power:} The most important functional feature is the range of supported data models.
This usually includes relational and popular NoSQL data models, 
such as JSON documents and property graphs, but can also include more specialized ones, such as spatio-temporal, 
or array-based models (S\ref{item: Sim_SupportSpatialFormats}, M\ref{item: Healthcare_MultipleDataFormats}). Furthermore, system properties such as availability (E\ref{item: ECom_Availability}, M\ref{item: Healthcare_Availability}), flexible consistency (E\ref{item: ECom_FlexibleConsistency}) and versioning (S\ref{item: Sim_ImmuntableData}) depend on the provided data models and systems.

\textbf{Query Expressiveness:} Almost equally important as the range of data models are the expressiveness and the variety of the provided query languages.
It has to be distinguished between \emph{data definition} (DDL) and \emph{data manipulation} (DML).
Concerning DDL, it is of interest whether or not the system supports the definition of an explicit schema and if integrity constraints, complex domains and proactive concepts (e.g. trigger) are provided.
%
One significant characteristic of a DML is whether the language supports complex \emph{set-based} queries, \emph{point access} to single objects, or allowing to \emph{navigate paths} between linked objects. Apart from classical relational operators, complex queries may include non-relational operators such as graph (E\ref{item: ECom_ComplexDataAnalysis}, M\ref{item: Healthcare_ComplexAnalysis}), array (M\ref{item: Healthcare_ComplexAnalysis}) or spatio-temporal ones (S\ref{item: Sim_ComplexAnalysis}).
Moreover, some query languages support recursive queries or user defined functions.

\textbf{Push-based Access:} While traditional systems focus on pull-based data access, newer systems, such as Meteor \cite{onlineMeteor2020}, also provide push-based queries which are particularly suitable for frequently changing data (S\ref{item: Sim_RealTimeAnalysis}, \ref{item: Healthcare_RealTime}).

\textbf{Data Security/Privacy:} For some applications (M\ref{item: Healthcare_Security_Privacy}), the support of special mechanisms for data security and privacy, such as encryption \cite{Salomon2003} and anonymization \cite{Wong2010} techniques, is essential and has to be supported.

\textbf{Flexible Adaption:} In order to support the aforementioned features, it is vital to enable the system to find an underlying data store landscape that meets the requirements formulated by the user. Thus, it is necessary that the user has the possibility to annotate individual parts of the data schema with special service level agreements (SLAs). A particular case where such annotations are indispensable is described in the property E\ref{item: ECom_FlexibleConsistency} of Section~\ref{sec:UseCaseECommerce}. In addition to providing annotations and since the underlying conditions, such as user requirements, workloads,
or data characteristics, can change over time, a polyglot data store must be able to adapt flexibly at runtime. There are three types of adaptation:

\begin{inparaenum}[(1)] \label{inpara:Adaption level}
	\item By changing the parameter setting and/or hardware configuration of a particular data store (e.g. in/decrease of buffer size or quorum-configs). For example, to achieve a higher read- or write-throughput (E\ref{item: ECom_ReadThroughput}, S\ref{item: Sim_WriteThroughput}), in which only a subset of data (E\ref{item: ECom_FlexibleConsistency}) is affected, to prevent constraint violations.
	\item By performing \emph{data or function shipping} between the underlying data stores and/or the mediator in order to execute domain-specific operations (S\ref{item: Sim_SupportSpatialFormats}, M\ref{item: Healthcare_MultipleDataFormats}, M\ref{item: Healthcare_ComplexAnalysis}) or to \emph{split} workload across the data stores to run operations concurrently (further details in Section \ref{sec:QueryProcessing}).
	\item By changing the system's store-topology (e.g. by starting and interconnecting a new MongoDB instance) and distributing a subset of data according to a well-known distribution model, such as master-slave in order to increase the read throughput (E\ref{item: ECom_ReadThroughput}) or a multi-master style to handle heavy write-workloads (S\ref{item: Sim_WriteThroughput}).
\end{inparaenum}

\textbf{Automatic Adaption:} 
To relieve users, the polyglot system should not only be flexible to adapt, 
but also execute necessary adaptions automatically.
The adaption of a system can be triggered by a wide variety of possible reasons. In contrast to manually triggered changes (\emph{offline}), systems can also react proactively (\emph{online}) by comparing 
the current workload against specified constraints (SLAs) \cite{sadalage2013nosql} or by using a prediction model \cite{vogt2018polypheny}.

\textbf{Transparency:} The complex processes of adaption and data (re)distribution should be hidden to users by providing \emph{location}, \emph{concurrency}, \emph{distribution} and \emph{mobility} transparency~\cite{tanenbaum2007distributed,ozsu2020distributedsystems}.
At the same time, physical and logical data independence should be provided. While physical independence is mainly guaranteed by the underlying stores, logical independence must be ensured by the polyglot system.

\subsection{Indirect Requirements}\label{sec:QueryProcessing}

The seven features described in the previous section
ensure that the major and unique selling point of polyglot data stores is their transparent and integrated access to multiple heterogeneous data stores without decreasing the overall query performance or -- better -- even increasing it. Finding the best query execution plan (QEP) in this setup, however, comes with a variety of challenges.
Many of these challenges are similar to those well-known from the area of information integration~\cite{doan2012principles}.
A big difference, however, is that in a polyglot data store, 
the data is not only read, but also written.
Another difference is that we usually have more control over the individual data stores and know exactly what data is stored in which store and in what form.

\textbf{Operator Placement:} First of all, the query has to be decomposed into sub-queries (query-splitting) that are executed in the underlying stores. This process is based on several conditions such as the query capabilities of these stores (e.g. a spatial join needs to be pushed down to a geo-database (S\ref{item: Sim_SupportSpatialFormats})) and the current distribution of data amongst them. Second, the system does not only have to be able to reorder the operations of a query but also have to decide which operations can be pushed down into an associated store. Furthermore, the system has to determine when to move data temporarily from one store to another (data-shipping) in order to perform an operation and if it is possible to replace the functionality of one store by a set of completely different operations of another store (function-shipping, query rewriting). Sometimes, even a permanent migration of data might be the best choice. In this case, it has to be decided when to trigger an expensive migration process and how to perform it (e.g. eager/lazy, online/offline)~\cite{StorlKS20}.
Currently, there exists multiple heuristics to address
the problem of optimal operator placement~\cite{cardellini2016optimal, kruse2020rheemix}.

\textbf{Cross-Store Joins:} As soon as all sub-queries return the associated results, it is inevitable to combine them in a higher-level system layer using the best suited operation (e.g. bind-join, hash-join, or skew-join~\cite{gupta2016cross}). This combination is a non-trivial process that requires some kind of internal data model such as the relational model, associative arrays, or a hybrid model (e.g. combining relational and JSON). Via the internal data model, it is possible to transform data residing in one store transitively into the model of any other underlying data store while preserving extensibility.

\textbf{Cost Models:} Due to the additional operations described above, the comparison of QEPs becomes more complex and has to be based on location as well as provided functionality. Dynamic as well as cost-based approaches that reduce, for instance, data movement or operations within higher-level layers can be used.

\textbf{Semantic Conflicts:} Apart from problems in planning and optimizing a query, 
querying data across different data stores
leads to a main integration question: 
How can semantic completeness be achieved by providing the user with the set of all operations available in the underlying stores?
Every data store comes potentially with a set of slightly different interpretation of an operation \cite{liu2015management}. This causes the problem in mapping these inconsistencies such as in relation to specific data types and missing values (e.g. the use of null values or the discrimination between null \cite{gottlob1988closed} and unknown or the exception process when data is not available or interpreted as default) so that it has to be considered if the overall data store should keep different semantics, use a predefined semantic or allow the user to define the applied semantic on a per query basis.

\section{System Types}\label{sec:System types}

\input{pictures/figure_system_types.tex}
Integrating different data types, models, and functions is a challenging research field. This section gives a general idea of the different kinds of systems that tackle these problems.

In their work, \citet{lu2019multi} presented two categories of systems. The \emph{multi-model databases} follow the approach to integrate varying data models into one system (e.g. ArangoDB or OrientDB), whereas \emph{multi-modal systems} organize data not by model but by domain and provide access to domain-specific data types such as speech, images, videos or handwritten text.

Another promising approach is to integrate several different SQL, NoSQL, NewSQL \cite{duggan2015bigdawg}, or Stream-Processing-Systems into a mediation system, e.g. the mediator-wrapper approach \cite{cluet1998your}. Underlying data models are not integrated into a single database engine but are handled by selecting suitable data stores if applicable. Based on the work of Tan et. al. \cite{tan2017enabling}, we distinguish between four different system types: Federated Systems (Fig.~\ref{subfig:Federated}), Polylingual Systems (Fig.~\ref{subfig:Polylingual}), Multistore Systems (Fig.~\ref{subfig:MultiStore}) and Polystore Systems (Fig.~\ref{subfig:PolyStore}). The systems we analyzed in this section were classified as one of these four architectures.

The system types differ in the composition of their underlying data stores and the number of query languages/interfaces they offer, as depicted in Figure~\ref{fig:SystemTypes}. Federated and Polylingual Systems use a homogeneous set of data stores beneath their mediation layer, whereas Multi- and Polystore systems rely on heterogeneous ones. Furthermore, Federated and Multistore Systems only offer one query language/interface, while Polylingual and Polystore Systems provide many interfaces.
In \cite{tan2017enabling}, the authors refer to the currently described Polylingual Systems as \emph{polyglot systems}. To avoid confusion with the term \emph{Polyglot Persistence}, which was already introduced by \citet{sadalage2013nosql}, we renamed the term \emph{polyglot systems} to \emph{polylingual systems}.

Poly- and Multistores provide transparent access to a set of (interconnected) data sources that reside in a static or dynamic topology. This can be accomplished by one or few query interfaces using a virtual global schema mapped to local schemas via views \cite{Halevy01, levy1996querying, garcia1997tsimmis, friedman1999navigational} or tuple-generating-dependencies \cite{doan2012principles}.

Furthermore, the architecture of these systems can be classified according to the CAP theorem \cite{gilbert2002cap}. Since they can choose and set up their underlying data stores according to the user's current needs, they can theoretically be CA, CP and AP at the same time (never all at once for the same application). Finally, a system's ability to extend to new data stores and provide a dedicated framework/API for integrating them is of utmost importance for its adaptability.

Two other solution families are similar to the \emph{polystore/multistore} concept in that heterogeneous stores are to be accessed. Despite this, they differ in the data sources' scope or the access mechanism. The so-called semantic data lake represents the first kind of solution. They provide uniform access to multiform data but do not consider any data movement across the underlying systems or duplication of data in multiple stores. They are simplifying the mediation to semantic data lakes, often utilizing \emph{Global-as-View} mapping and technologies from the semantic web.

The second kind of system is represented by the solutions mapping relational databases to RDF \cite{spanos2012bringing}, and Ontology-Based Data Access over relational databases \cite{wimalasuriya2010ontology}, such as Stardog, Ontop, Morph, Ultrawrap, Mastro. These solutions do not consider querying large-scale data sources, e.g., HDFS or NoSQL stores, but integrating multi-variety data from decoupled and remote sources, e.g., from OpenData portals.

The most significant difference to the \emph{polystore/multistores} considered in this paper is the task to find which store or combination of stores answers best a given query (transactional, analytical or hybrid) or which store to move all/part of the data to.

\section{Analyzed Systems}\label{sec:Analyzed Systems} 
    
This section introduces a set of different and representative polyglot managed data stores with their functionalities and special features. We selected these systems for their representative Poly-/Multistore approach and general availability, focusing on a differentiated perspective to support multiple use cases. Due to the variety of systems, this survey can provide extensive details of the concepts in terms of query interfaces, query planning, execution, and migration. Other known systems such as \cite{wang2017myria}, HadoopDB \cite{abouzeid2009hadoopdb} and SparkSQL\cite{armbrust2015spark} are beyond the scope of this paper. Table~\ref{tab:ClassificationCurrentSystems} shows an the considered systems and their classification as \emph{Multi-} and \emph{Polystore}, respectively. \emph{Loosely-coupled} systems correspond to a network of often autonomous data stores where the mediator has read rights but no rights to write data or reconfigure the individual stores, able to migrate data sets between them. In general, they bear a strong resemblance to virtual integration systems.
In contrast, \emph{tightly-coupled} systems were often designed from the beginning as one overall system, capable of applying writes in order to combine multiple processing features instead of simply performing integration of stored data into a common global schema. This leads to a set of individual stores for which data migration is possible and whose individual stores are typically not autonomous so that the mediator can adapt the configurations to actual requirements. Hybrid systems are those in which some stores are tightly coupled, and others are loosely coupled. 

\begin{table}
\centering

\begin{tabular}{p{2.0cm}|p{3.3cm}|p{3.3cm}} 
& \multicolumn{1}{c|}{\emph{multistore}} & \multicolumn{1}{c}{\emph{polystore}}   \\\hline
\multirow{10}{*}{\emph{loosly-}\emph{coupled}} &  \textbf{PolyBase}~\cite{dewitt2013split}& \textbf{Myria}~\cite{wang2017myria}\\
 & \textbf{BigIntegrator}~\cite{zhu2011querying}  & \\
 & \textbf{FORWARD}~\cite{ong2014sql++} & \\ 
 & Apache Drill~\cite{onlineApacheDrill2020} & \\
 & Apache Calcite~\cite{begoli2018apache} & \\
 & TATOOINE~\cite{bonaque2016mixed} & \\
 & QoX~\cite{simitsis2012optimizing} & \\
 & QUEPA~\cite{maccioni2016quepa} & \\
 & Odyssey~\cite{hacigumucs2013odyssey} & \\
 & DBMS+ (Cyclops)~\cite{lim2013fit} & \\\hline
 \multirow{4}{*}{\emph{tightly-coupled}} & \textbf{RHEEM}~\cite{agrawal2016rheem} &  \textbf{ESTOCADA}~\cite{alotaibi2019towards}\\
& \textbf{MuSQLE}~\cite{giannakouris2016musqle} & \textbf{Polypheny-DB}~\cite{vogt2018polypheny}\\
& AWESOME~\cite{dasgupta2016analytics, dasgupta2017generating}  & \\
& \textit{HadoopDB}~\cite{abouzeid2009hadoopdb}  & \\\hline
\multirow{2}{*}{\emph{hybrid}} & \textbf{CloudMdsQL}~\cite{kolev2016CloudMdsQL}  & \textbf{BigDAWG}~\cite{duggan2015bigdawg}  \\
& \textit{SparkSQL}~\cite{armbrust2015spark} & \\ 
\multicolumn{3}{c}{}\\[-0.5em]
\end{tabular}
\caption{Overview of existing systems for polyglot data management classified as \emph{multi-} or \emph{polystore} tightly utilizing systems for processing or as data integration solution. The stores we discuss and compare in detail are in bold.
The italicized systems are out of the scope of this paper and were not discussed further.}
\label{tab:ClassificationCurrentSystems}
\vspace*{-0.6cm}
\end{table}

\subsection{CloudMdsQL}\label{sec:CloudMdsQL}

\begin{figure}
    \centering
    \includegraphics[width=\textwidth]{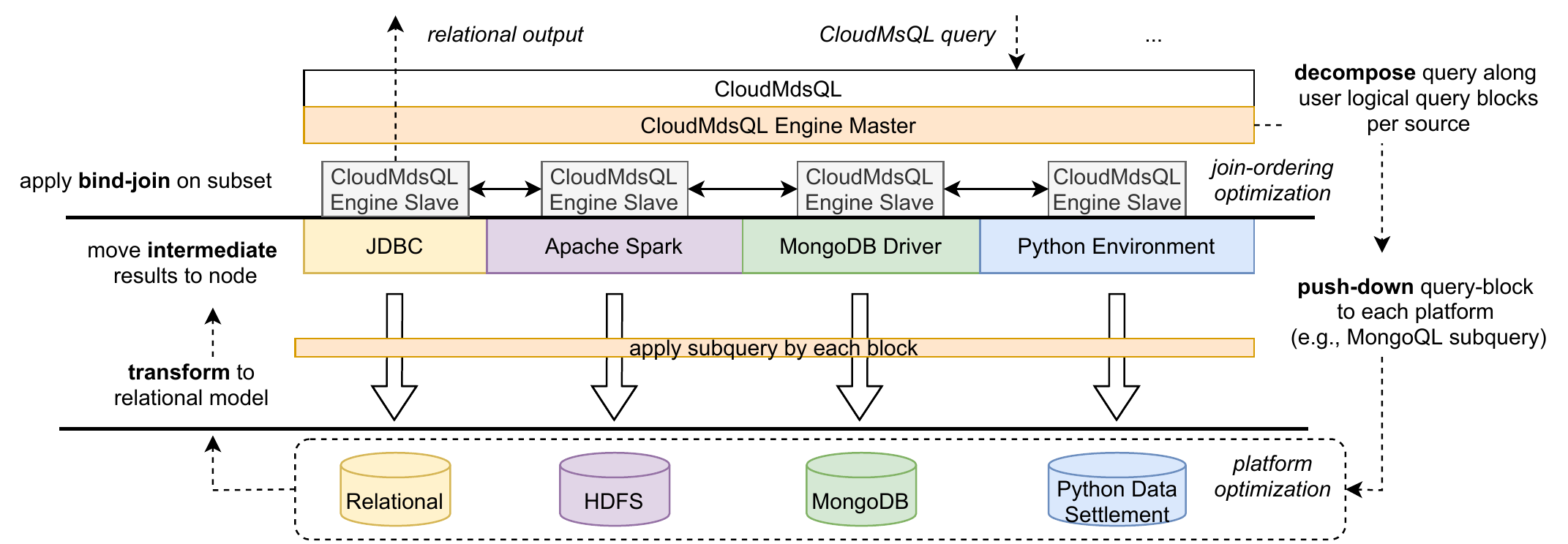}
    \caption{Conceptual architecture of CloudMdsQL, adopted from \cite{kolev2016CloudMdsQL}}
    \label{fig:CloudMdsQLConceptual}
\end{figure}

The \emph{CloudMdsQL} multistore system\footnote{Cloud Multidatastore Query Language} \cite{kolev2016CloudMdsQL} aims at providing a powerful functional SQL-like language as part of the \emph{CoherentPaaS} project \cite{onlineCoherentpaas2020} and implemented in the LeanXScale system \cite{kranas2021parallel}. The \emph{CloudMdsQL} system has been developed as an abstraction layer to retrieve data from different stores, keeping the underlying store and query semantics. The system compiles queries into a relational query framework, in which each sub-queries contains the native database query to select the source data. LeanXScale executes CloudMdsQL query distributed, transforming the result of sub-query into an intermediate table, partitioning them to apply a distributed processing. Figure \ref{fig:CloudMdsQLConceptual} show the conceptual design of the CloudMdsQL system. The input is a pre-formulated CloudMdsQL query, in which each embedded block already describes the decomposition to access the correct data store. Nodes responsible for the accessed data store execute each native subquery (e.g. user-defined MongoQL query to retrieve documents in MongoDb). The system merges partial results by transforming each result into an intermediate relational format, joining each transformed result from sub-queries by utilising a \emph{bind-join}. 

CloudMdsQL provides a familiar query interface to achieve semantic completeness of all associated stores by defining sub-queries with the native data store language (e.g. MongoDB statements or SQL). The base language is SQL oriented and additionally provides Python as an integrated connector to retrieve or transform queried results as an intermediate step in the plan. The base language contains named table expressions, which are wrapped into functional calls for the wrapper, translating calls to the native API of the wrapper without an intermediate integration step. \emph{CloudMdsQL} provides a minimal set of types in order to ensure the most used types in data stores. Including scalar (integer, float, string, binary and timestamps) composite types (array and associative array), complemented by arithmetic, concatenation and access operations.

The language itself is much more restrictive, making the language more practical. When using different native query languages, these sub-queries have to be defined in the context of a so call table named expression, transforming the result into an intermediate table format. In contrast to these native queries, \emph{CloudMdsql} provides a language extension to incorporate distributed processing systems such as Apache Spark. The extension provides map-, filter- and reduce (MFR) operators applicable for big-data ad-hoc queries. MFR statements consist of a sequence of operating instructions in which the system incorporates datasets by transforming results into a Resilient Distributed Dataset (RDD as the Apache Spark working unit). Each of the MFR operations applies a transformation on the tuples of the dataset and produces a new dataset for subsequently part of the complete \emph{CloudMdsQL} query.

The system applies two specific optimisations for optimal execution of all these different operations and all these blocks of native queries. For efficient merge of two results \textit{A} and \textit{B} from sub-queries, the system applies a bind-join reformulation. In contrast to an equality \textit{JOIN ON} operation, the bind-join allows retrieving data from heterogeneous sources, as long as the acquired source provides a filtering technique. Therefore, it executes and loads one side \textit{B} thoroughly to subsequently select a matching partner by filtering for a key relationship condition on side \textit{A}. However, since the complete side B have to be loaded first, the system estimates the cardinality of each result by using the cost model of the native's store when available. Otherwise, when no cost models are available, the system looks for exceeding a threshold of join keys to fall back to a hash-join. The other optimisation considers the MFR rewriting to bring filter operations in MFR statements forward as much as possible. Therefore, the query planner controls the planning by three rules: \begin{inparaenum} \item A name substitution replaces column names in predicates of MFR's filter operations with the data reference of the MFR mapping. \item A \emph{reduce-filter} switch rule changes the order of \texttt{reduce} and \texttt{filter} operations to shrink data object for a reduction tasks, whereas \item a map-filter switch rule change the order for \texttt{map} operations analogously.\end{inparaenum} Each of these rules are applied in the scope of the MFR statements for the execution.

To execute \emph{CloudMdsQL} queries, a query engine called LeanXScale distributed the processing work within a master-worker environment. LeanXScale is a Hybrid Transactional Analytical Processing, bringing OLAP and OLTP together and consisting of a planning controller node and multiple workers, directly linked to exchange query plans and other data. Worker nodes execute the query, and controller nodes plan the query. Generated execution plans are split into multiple sub-plans, each assigned to a respective worker node, collocating the underlying data store and exchanging results. Since the CloudMdsQL approach converts sub-queries into intermediate tables, relational operations are available, improving the runtime of LeanXScale by partitioning the base tables for the scan operations among all worker nodes. The partitioning step respects stateful operations such as group-by and reduce.

\begin{figure}
    \centering
    \includegraphics[width=\textwidth]{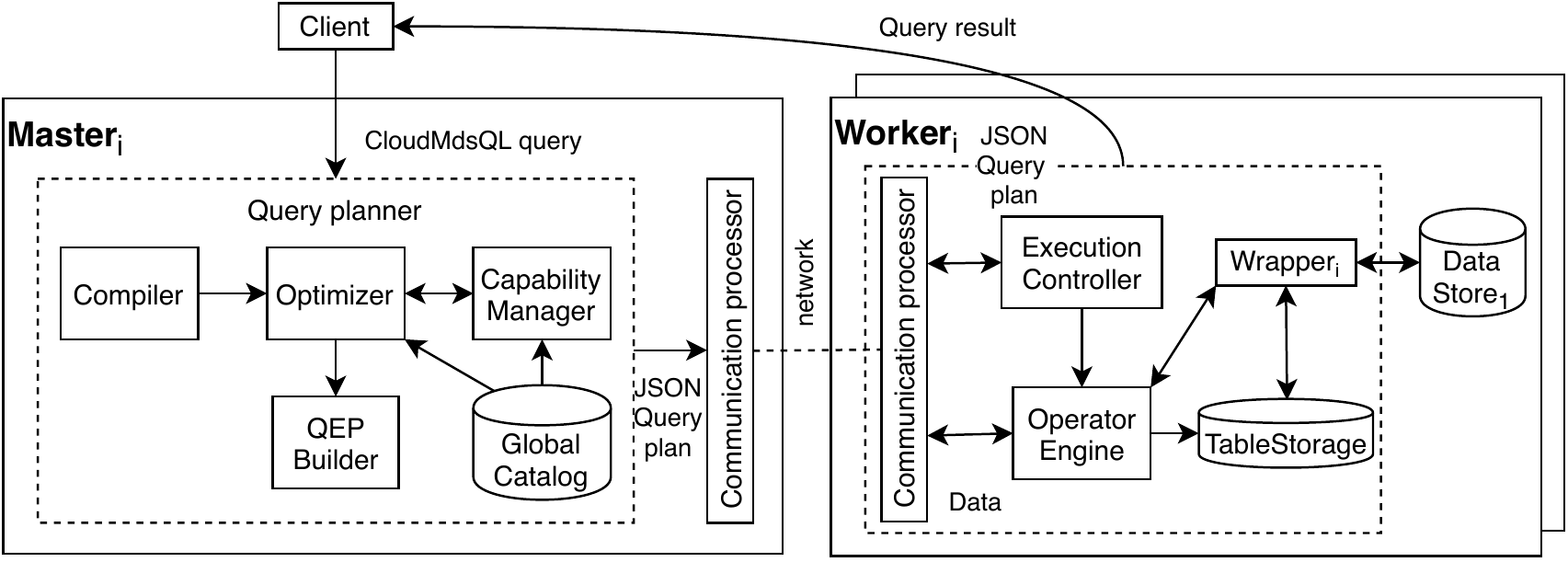}
    \caption{Distributed architecture of the CloudMdsQL engine from \cite{kolev2016CloudMdsQL}}
    \label{fig:CloudMdsQL}
\end{figure}

Planning a query by a controller node produces a JSON-based operator plan transferable to assigned nodes. The plan is an acyclic graph consisting of relational operators in which the leave nodes are named table result references coming from the native sub-queries. These table results can be referenced anywhere in the plan and used as input for other operators. The uppermost operator selects the first worker node to use. Figure~\ref{fig:CloudMdsQL} shows the technical design for this distributed execution. 

Cost model the bind-join are collected by keeping database-specific features under API contract in the wrapper, besides periodically updated database statistics and estimations of sub-queries. CloudMdsQL considers reorderings and improvements in the scope of these subsets of operations and not native queries. As to the lack of cost models for some systems, the \emph{CloudMdsQL} engine also allows user-defined cost models and default costs when nothing can be collected.

\emph{CloudMdsQL} is used in the CoherentPaaS \cite{onlineCoherentpaas2020} platform to retrieve and process data from heterogeneous data stores for analytical purposes and can be used in conjunction with the distributed LeanXScale query engine, accepting \emph{CloudMdsQL} queries. Evaluations were made in \cite{kranas2021parallel} showing a significant impact of partitioning results onto multiple workers and merging them back using the bind-join approach.

\subsection{BigIntegrator}

The focus of the \emph{BigIntegrator} system \cite{zhu2011querying} is to access relational DBMS and BigTable databases, especially cloud-based NoSQL stores. To achieve this, \emph{BigIntegrator} introduces a query interface supporting a SQL-like query language. Incoming queries and generated QEPs contain relational operations with SQL and BigTable database function calls of GQL\footnote{Google BigTable Query Language}. Therefore, GQL represents a subset of SQL and sacrifices the full expressiveness of SQL in favour of scalability since only basic filter (selection) predicates are supported. To compensate these restrictions, more complex operations as \emph{joins} or the \emph{like} operator are provided by \emph{BigIntegrator}. BigIntegrator tries to \emph{pushing down} as many operations as possible, handled by corresponding stores in the post-query step whose remaining not supported operations are fulfilled by the query language-specific \emph{absorber} plugins, as part of each wrapper. The query engine accesses each \emph{absorber} as a plugin in the wrapper and replaces the source with the respective store access. This access results in a QEP, containing both relational operators and calls to the absorber plugin (\emph{API-calling} module), in which the system moves as many store-specific filter predicates into the plugin implementation. In the end, a \emph{finaliser} plugin receives the plans and executes them. In contrast to other systems such as \emph{BigDAWG}, BigIntegrator only works with the BigTable systems and cannot be integrated with other NoSQL systems or models.

BigIntegrator is designating a plugin architecture with expressiveness using GQL for a more straightforward integration of different data stores. A new system needs to provide or use an existing \emph{absorber}-  and \emph{finaliser} plugin to grant access to query capabilities of the underlying store. Therefore, BigIntegrator transforms the input query into a Datalog \cite{eiter1997disjunctive} query, which can contain both source predicates and non-source predicates (NSPs). The absorber manager takes the Datalog query, which calls and, for each source predicate referenced in the query, calls the corresponding absorber of its wrapper. The manager collects referencing source predicates and replaces them with the native access filters. This abstraction allows decoupled filters and other predicates to produce an algebra expression containing the access to stores and NSPs. Each access is passed toward the call on the corresponding finaliser of its wrapper, which transforms the access into interface function calls.

\subsection{ESTOCADA}

\begin{figure}
    \centering
    \includegraphics[width=\textwidth]{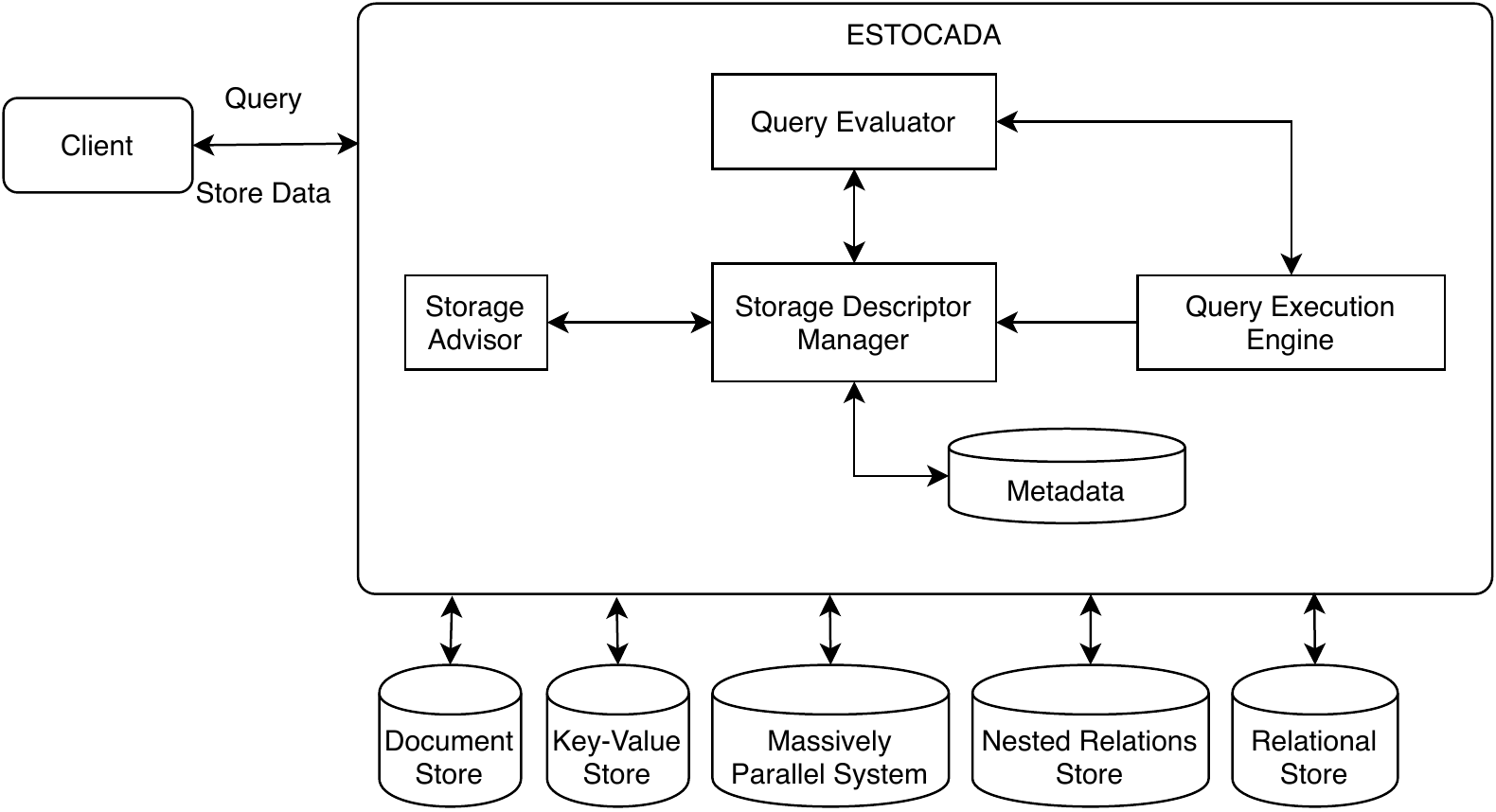}
    \caption{The architecture of the ESTOCADA system from \cite{alotaibi2019towards}}
    \label{fig:estocada}
\end{figure}

\emph{ESTOCADA} \cite{alotaibi2019towards} is a multistore system to optimise the performance of applications by reducing the cross model problem to a single model problem. It queries data fragments across heterogeneous stores and uses view-based query rewriting with a decoupled result integration. 

ESTOCADA supports data in the format as JSON, XML, key-value, graph and nested relations or full-text. Fragments of datasets can be consists of multiple data models and are transformed into an internal pivot model to leverage development and query processing. The internal pivot model consists of a relational model expanded by constraints, specifying the application related data model. It allows representing relational queries expressed in multiple supported languages, splitting a single query into multiple ones using a query tree block approach similar to \ref{sec:CloudMdsQL}. Combinations of subqueries are formulated by an own query interface called  \emph{QBT\textsuperscript{XM}}, in which each matching native query language is expressed in an individual tree block.

The architecture of ESTOCADA consists of two main distinct components. These can be any NoSQL store, a key-value store, a document store, one for nested relations or a relational one. The Storage Advisor module splits each data set into fragments and chooses the backend system for each fragment based on the underlying data model, where to store them accordingly. ESTOCADA decides fragmentation by the amount of application workload, driven according to the exact amount of data accessed as parameterised access pattern or via a heuristic on top of the underlying data model.

The mapping of data to stores is handled through a storage descriptor. These descriptors specify what is contained in the fragment and where it can be found. The containment is defined by the query in the target primary data model, where the source is designated by given a source schema for the database where the outgoing data set is stored. Thus wrapper functionality and mapping information are saved for each fragment, decoupling the system from using extra wrapper components for each tied underlying store. This approach allows the integration of new systems by providing individual store descriptors, respecting the constraints of the new system.

The query interface of this system provides an integration language, coined \emph{QBT\textsuperscript{XM}}, which is based on the Query Block Trees of System R and provides the properties to express each block in a different data store's matching query language and data model. To achieve the best possible performance from the available data stores, \emph{ESTOCADA} automatically distributes and partitions the data across the different data stores, which are entirely under its control and hence do not have any autonomy. Therefore, it is a tightly-coupled multistore system.

Reformulation of relational queries (pivot) on views for each data store, holding constraints. ESTOCADA solves the reformulation problem using the Provenance-Aware C\&B algorithm, respecting conditions in relational algebra.

ESTOCADAs generic model approach and their architecture are motivated by large-scale e-Commerce and open data warehousing systems for digital cities, dealing with the requirements in Section~\ref{sec:UseCaseECommerce}.

\subsection{Myria}\label{sys:myria}

Myria \cite{wang2017myria} is a federated data-analytics system with an imperative-declarative hybrid language called \emph{MyriaL} and optional support for Datalog \cite{eiter1997disjunctive}. It facilitates the expression of complex data analytic tasks and wraps each declarative statement with imperative constructs, such as variable assignments and iterations. Myria implements support for user-defined functions and aggregates exposed by a Python API for full acceptance. Myria uses shared-nothing architecture and enhances the query engine (MyriaX) with iterative processing focusing on horizontal elasticity with scale-in and scale-out. It provides a data-ingestion for HDFS and multiple cloud storage, using YARN container \cite{vavilapalli2013apache} in a distributed setting.

\begin{figure}
    \centering
    \includegraphics[width=\textwidth]{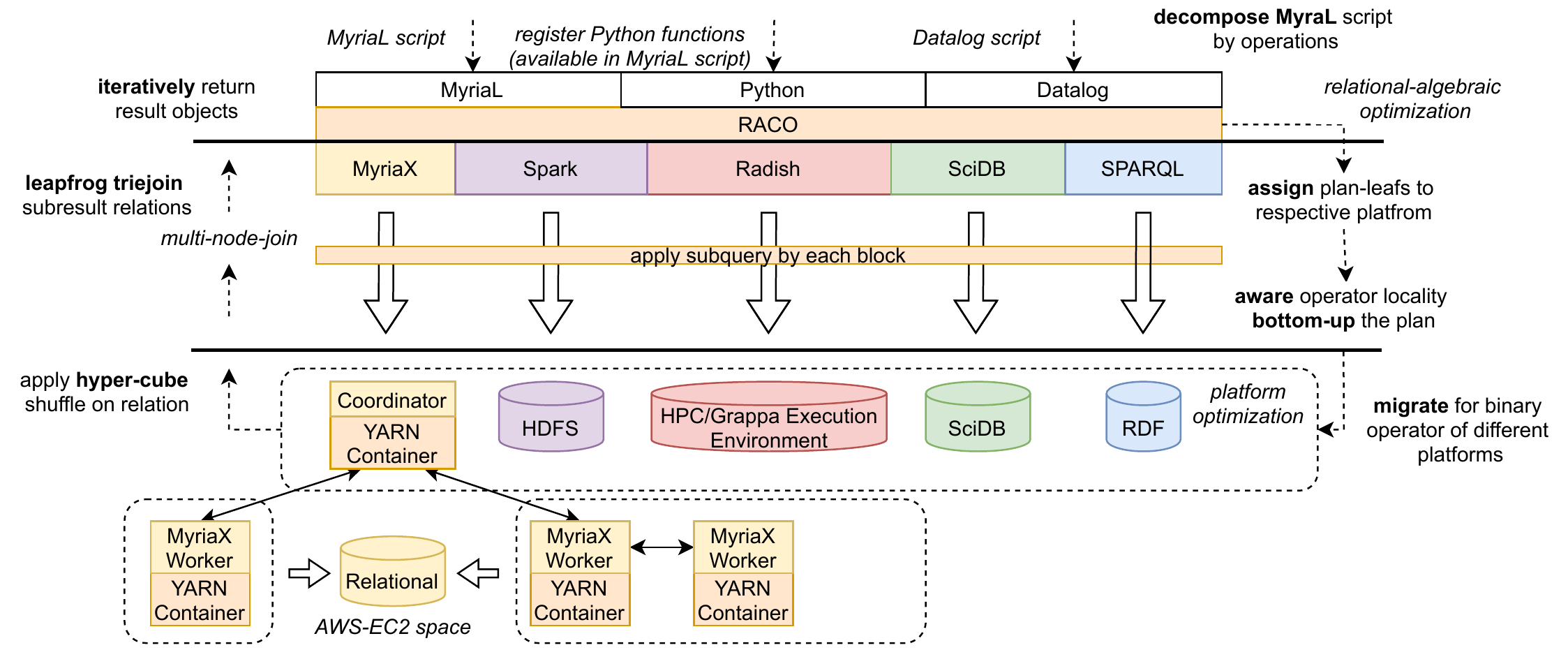}
    \caption{Conceptual architecture of the Myria Big-Data Management System, adopted from \cite{wang2017myria}}
    \label{fig:myria}
\end{figure}

Myria uses a relational-algebraic optimisation with a dedicated relational algebra compiler called (RACO). RACO is a relational, local-aware, algebraic optimiser using a rule-based approach and extends it by imperative constructs to capture the semantics of array, graph and key-value data. Figure \ref{fig:myria} shows the architecture containing the outer view top-down. Inputs in MyriaL processed by RACO can produce federated execution plans considering computation and data movement across the subsystems. Extensibility of to new backend system can be provided by mapping the relational operators to the new API, AST or supported query language concepts. The database administrators must implement their functions to retrieve metadata for newly coupled backend systems. Therefore the new system needs to provide rewriting rules for the RACO compiler.

RACO extends the model in multiple ways. It enables iterative processing, which is a recurrent requirement in performing machine-learning- and graph-analytics task. Therefore the RACO supports a \emph{do-while} loop, in which the content is executed each time and checking a loop boundary. The termination conditions is synchronized by a relational sub-query, those result contains exactly \emph{one} tuple with a \emph{boolean} attribute. The other loop extension is a \emph{do-until} loop, in which asynchronous processes are enabled. Moreover, RACO integrates a flat map operator \cite{chintapalli2016benchmarking}, in which non-1NF values are formed into multiple ones and the support of a so-called \emph{stateful apply} operator. This \emph{stateful apply} is used to provide window function, making sliding window operations on datasets available.

Myria's query execution is based on RACO and translates the inputs formulated by MyriaL into a logical algebra. Query plans are in the form of \emph{operator-graphs} and can contain cycles for the iterative extensions. A query consists of multiple individual query fragments, in which a subset of operators within the graph are grouped and executed by an individual shared-worker thread. The query federation assigns each leaf of the operator tree to those platforms where the data exist. Minimizing data movement is an open challenge and considered by integrating a data-movement operator into the plan when data lies on a different platform for binary operation.

This operator set consists of mentioned loop extensions and the known relational operators such as aggregates and joins. Especially joins are supported in MyriaX by using \emph{HyperCube} \cite{beame2017communication} and \emph{Shares} \cite{afrati2010optimizing} data distribution algorithm, which is leveraged in a decoupled operator called \emph{Tributary Join}. These multi-joint algorithm tries to process the conjunctive query in one communication round within the distributed setting. It builds a balance of joins among the servers to avail from their parallel execution and reduces the amount of additional data-movement operations across engine boundaries.

Intermediate query results are exported as CSV tabular data and stored in HDFS. This file approach offers higher connectivity for external systems but results in lower transfer performance when coupled engine provides do not provide task-optimized imports. Since the exponential growth of implementing custom operators for each engine combination, MyriaX provides a data movement component called PipeGen \cite{haynes2016pipegen} in order to facilitate the import and export functions for CSV. PipeGen is an extension for DBMS to speed up the import and export, using Apache Arrow \cite{onlineApacheArrow2020} with their network protocol and compressed message format.

The Myria big-data management project is successfully used in multiple domains, such as natural language processing, neuroscience, astronomy and oceanography.

\subsection{Polypheny-DB}

\emph{Polypheny-DB} is a research system designed by the database group at the University of Basel from a vision in \cite{vogt2018polypheny} towards a first version to support hybrid transactional and analytical processing (HTAP) workload, a combination of OLTP and OLAP requirements. The project aims to create a self-adaptive polystore (in a cloud environment) that provides access to a heterogeneous set of data while incorporating the characteristics of the system's workload to balance out and improve data distribution and underlying data stores choices. In contrast to other systems presented and discussed in this paper thus far, \emph{Polypheny-DB} is - in its current state - a vision and not a fully functional system \cite{vogt2018polypheny}. Nonetheless, several smaller projects exist in the research group's portfolio intended to fulfil the roles of components of the \emph{Polypheny-DB} system. In the following, we present the vision of \emph{Polypheny-DB} and the components the developers want to use in their endeavour.

\begin{figure}
\centering
    \includegraphics[width=0.8\textwidth]{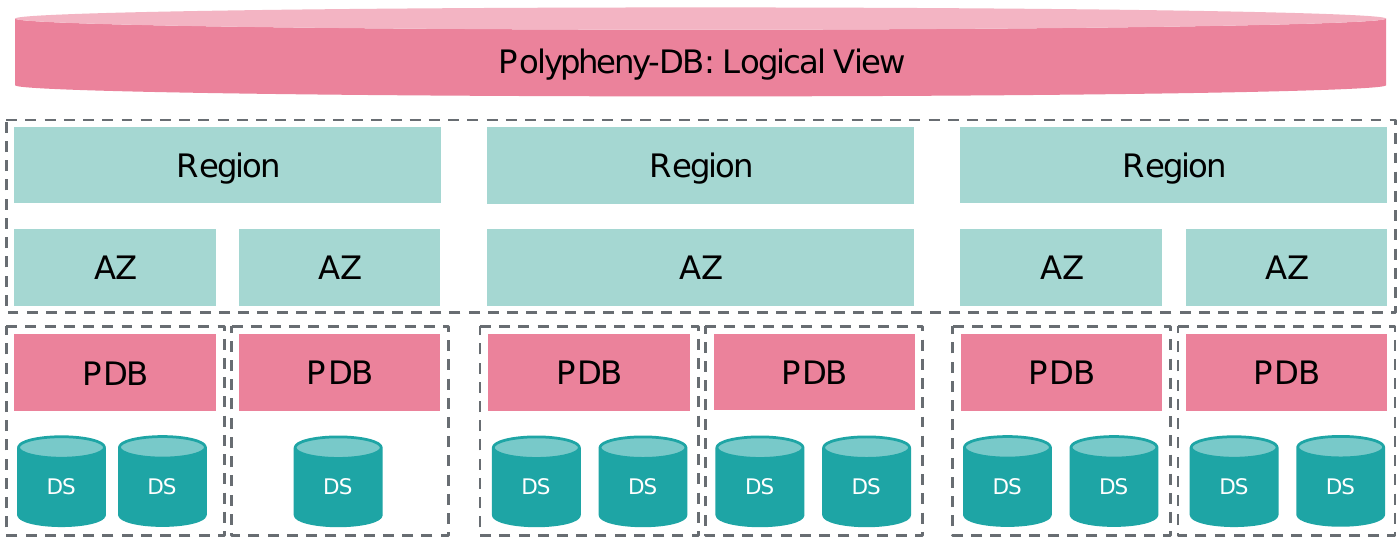}
    \caption{\emph{Polypheny-DB's} layered architecture consisting of Regions, Availability Zones (AZ) and \emph{Polypheny-DB} instances with their underlying datastores \cite{vogt2018polypheny}}
    \label{fig:polyphenydb_layer}
\end{figure}

A major design decision for \emph{Polypheny-DB} is its distinction between two data management layers (Figure~\ref{fig:polyphenydb_layer}):
\begin{inparaenum}[(1)]
    \item A global data partitioning and replication layer in the Cloud, which is based on user requirements (e.g. consistency level or availability quota) and the resource optimization of Cloud providers.
    \item A local level with polystores located in individual Cloud data centres. These stores are envisioned to leverage the strengths and sweet spots of different data models, data stores, and storage media.
\end{inparaenum}

The data partitioning is supposed to be handled
\begin{inparaenum}[(1)]
    \item explicitly by incorporating user-defined flags on attributes or 
    \item implicitly by analysing the system's workloads.
\end{inparaenum} 
The partitioning is built on previous work on Cumulus~\cite{fetai2015cumulus}, supposed to be handled by QuAD \cite{fetai2017quad}, an adaptive quorum-based replication protocol that dynamically selects an optimal quorum configuration based on sites' load and network latency. While QuAD can adapt its quorum strategies by considering the site's properties and therefore outperforms static approaches, it cannot adapt to user-defined requirements (SLAs) or the system's actual workload.

Polypheny-DB utilises data partitioning and replication by employing a cost model derived from an extended version BEOWULF~\cite{stiemer2016beowulf} cost model. The extensions include the specification of user-defined requirements as service-level objectives (SLO)s, compiled from SLA agreements. Regarding the translation from SLAs to SLOs, \emph{Polypheny-DB's} developer refers to PolarDBMS, which included an approach for this \cite{fetai2014polardbms}.

\emph{Polypheny-DB}s query interface incorporates multiple query languages and data models. Besides a CRUD interface to point-access, it includes at least a minimal SQL dialect and OpenCypher, an open query language for property graphs, built upon subsystem Icarus \cite{vogt2017icarus}. In \cite{vogt2018polypheny}, the authors aim to extend Icarus's ability to connect to different SQL based, relational databases by adding support for document and graph stores. For the internal data representation, incoming queries are translated into an algebraic tree to use mathematical models unifying relations, graphs and matrices like associative arrays. In order for \emph{Polypheny-DB} to provide ACID guarantees for transactions, the system will use a two-phase commit protocol for coordination, knowing the danger of deadlocking in case of failures.

On the local polystore level \emph{Polypheny-DB} dynamically distribute and replicate data to add or remove datastores if needed. Besides the difficulties in dealing with the automatic management of stores, a sound way to estimate current and future workloads have to be implemented to change the system accordingly. For this, workload predictions from Cumulus \city{fetai2015cumulus} and BEOWULF \cite{stiemer2016beowulf} are intended to be incorporated. For its query planning needs, the system builds on efforts done on Icarus \cite{vogt2017icarus}. Icarus's query planning component estimates the execution time of an incoming query \begin{inparaenum}[(1)]\item by analysing its used operations \item and comparing its structure, entities, functions and operators to similar queries already encountered by the system. \end{inparaenum}
The estimation defines an upper time limit for further analysis and the generation of queries execution plans\footnote{Methods for generating the plans include: query-splitting, and execution on different datastores, data migration}, which are assigned a cost. The plan with the lowest cost is then given to the query execution engine.

In addition to the polystore functionalities, the developers of \emph{Polypheny-DB} want to include additional functionalities like temporal data management based on their system ARCTIC~\cite{brinkmann2015arctic}, an index structure for searching versions of data items by archiving queries, and multimedia retrieval provided by their system ADAM$_{pro}$ \cite{giangreco2016adampro} in conjunction with the multimedia retrieval engine Vitrivr \cite{rosetta2016vitrivr}.

Conclusively, \emph{Polypheny-DB} is one of the few systems aiming at a self-adaptive ability but is very vague in how that is to be achieved. The vision paper's primary component is the cost model, which determines almost every decision the system has to make. Considering all the variables and inputs (SLAs/SLOs, current/future workloads, per-Transaction overwrites, global and local data partitioning/replication), the cost function's scope becomes quite ambitious and poses one of the most significant difficulties. Additionally, the amount of different systems which are supposed to be included as components within \emph{Polypheny-DB} is very sizable and can be a problem on its own if everything is combined into one. In addition, a system like Icarus has to be extended and cannot be incorporated as-is.

\subsection{PolyBase}

\begin{figure}
\centering
    \includegraphics[width=\textwidth]{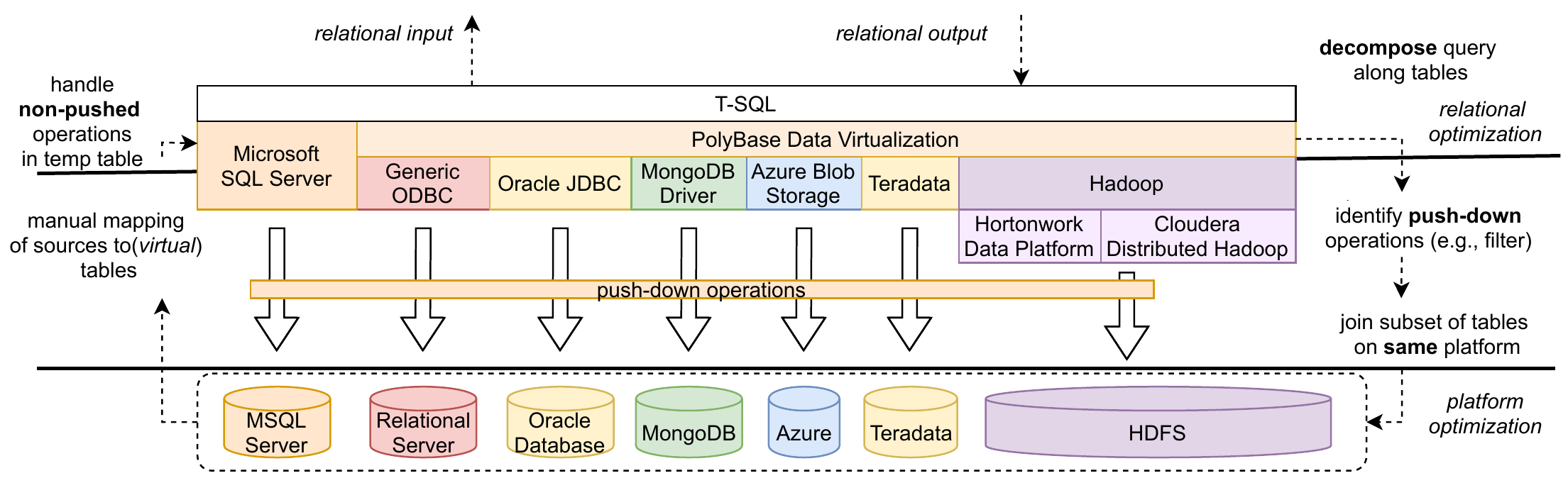}
    \caption{\emph{PolyBase} conceptual architecture adopted from \cite{dewitt2013split}}
    \label{fig:PoylBaseConcept}
\end{figure}

As part of Microsoft's efforts to improve data virtualisation for SQL Server, Microsoft introduces \emph{PolyBase} \cite{dewitt2013split} as an active component to retrieve data from a remote source and to shift access workload into an acquired system of the polyglot setup. Figure~\ref{fig:PoylBaseConcept} shows the conceptual design of the data virtualization. Users can query data directly from SQL Server, MongoDB, Teradata, Oracle, Hadoop clusters, and Cosmos DB using T-SQL without installing special client connection software. The system uses a generic Open Database Connectivity (ODBC) standard to connect to additional providers via third-party ODBC drivers. The main approach is to keep different data sets in a relational format. Therefore, the user must provide a mapping via tables from the inner PolyBase scope to the external data provider. For example, if data from MongoDB is to be queried using T-SQL, a so-called external table must be created manually in PolyBase, describing the mapping of the documents (possibly hierarchical) structure to the relational format. The manual mapping into columns with corresponding types must be explicitly defined for fields from the documents. The main goal of PolyBase is to provide a solution to two use cases:
\begin{inparaenum}[(1)]
    \item PDW requires data from Hadoop and returns the result to the user/application or
    \item PDW requires data from Hadoop and materialises the result as an output file in Hadoop for later usage by PDW or MapReduce.
\end{inparaenum}

\begin{figure}
    \centering
        \subfloat[PolyBase's Use cases for Hadoop/SQL interaction]{
            \includegraphics[width=0.45\textwidth]{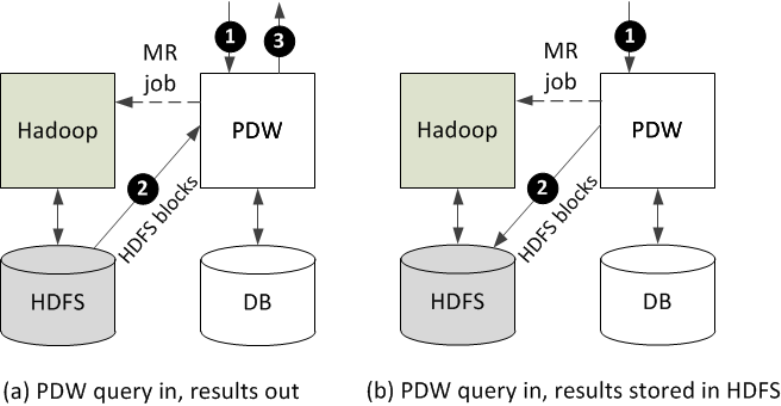}
            \label{subfig:PolyBase_usecases}
        }
        \hfill
        \subfloat[PDW architecture overview]{
            \includegraphics[width=0.45\textwidth]{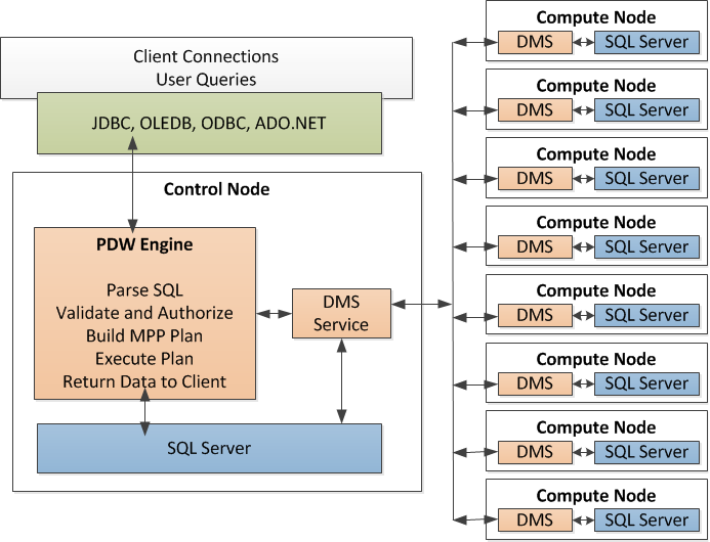}
            \label{subfig:PolyBase_pdw}
        }
        \hfill
        \subfloat[Structure of a single PDW Compute Node with embedded HDFS Bridge]{
            \includegraphics[width=0.45\textwidth]{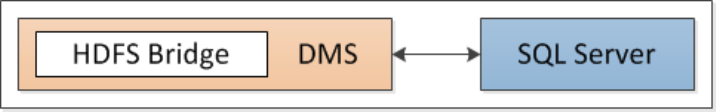}
            \label{subfig:PolyBase_computenode}
        }
        \hfill
        \subfloat[Detailed HDFS Bridges structure]{
            \includegraphics[width=0.45\textwidth]{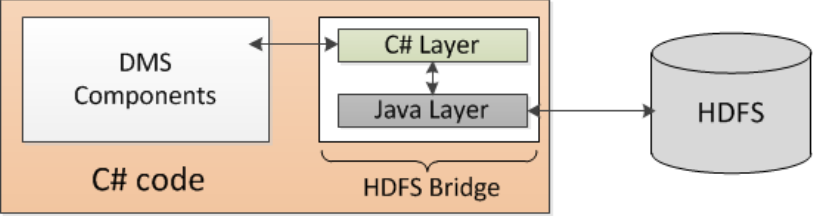}
            \label{subfig:PolyBase_hdfsbridge}
        }
    \caption{The architecture overview of PolyBase as part of Microsoft's SQL Server PDW v2 \cite{dewitt2013split}}
    \label{fig:PolyBase}
\end{figure}

PDW basic architecture comprises a control node and a variable amount of worker nodes. Figure~\ref{subfig:PolyBase_pdw} show the technical design of PolyBase with its distinct PDW and interceptor component append to the SQL server.

For this task, HDFS data can be referenced in \emph{PolyBase} as external tables, which makes the correspondence with the HDFS file on the Hadoop cluster possible. By providing an external relational \emph{view} as well as their current execution costs, PDW allows manipulating both the native and external tables expressed by location-transparent SQL queries over a GAV schema. These are executed according to a distributed SQL execution plan, in which the query optimiser decides to \emph{split} the query and to push SQL operators to the external store by building jobs for projection and selection on the external tables or when asymmetric join is applied on two external tables. 

In many cases, PolyBase can simplify the pushdown of the JOIN operation to improve query performance. When a join is performed on an external source, this reduces the amount of data movement and increases query performance. Without JOIN available, the data to be linked is moved locally from the tables to a temporary data source and only then linked. Joins, projections, aggregations, filters, and statistics determinations are forwarded to the data source via the connection using ODBC. For Hadoop Cluster and MongoDB, on the other hand, filters and statistics calculations are only partially available, while JOINS are not possible with them.

The overall objective is to minimise the data transfer amount between HDFS and PDW. Data imported/exported to/from PDW is processed in parallel and integrated into the same PDW service that shuffles PDW data among compute nodes.

\subsection{BigDAWG}

\begin{figure}
    \centering
    \includegraphics[width=\textwidth]{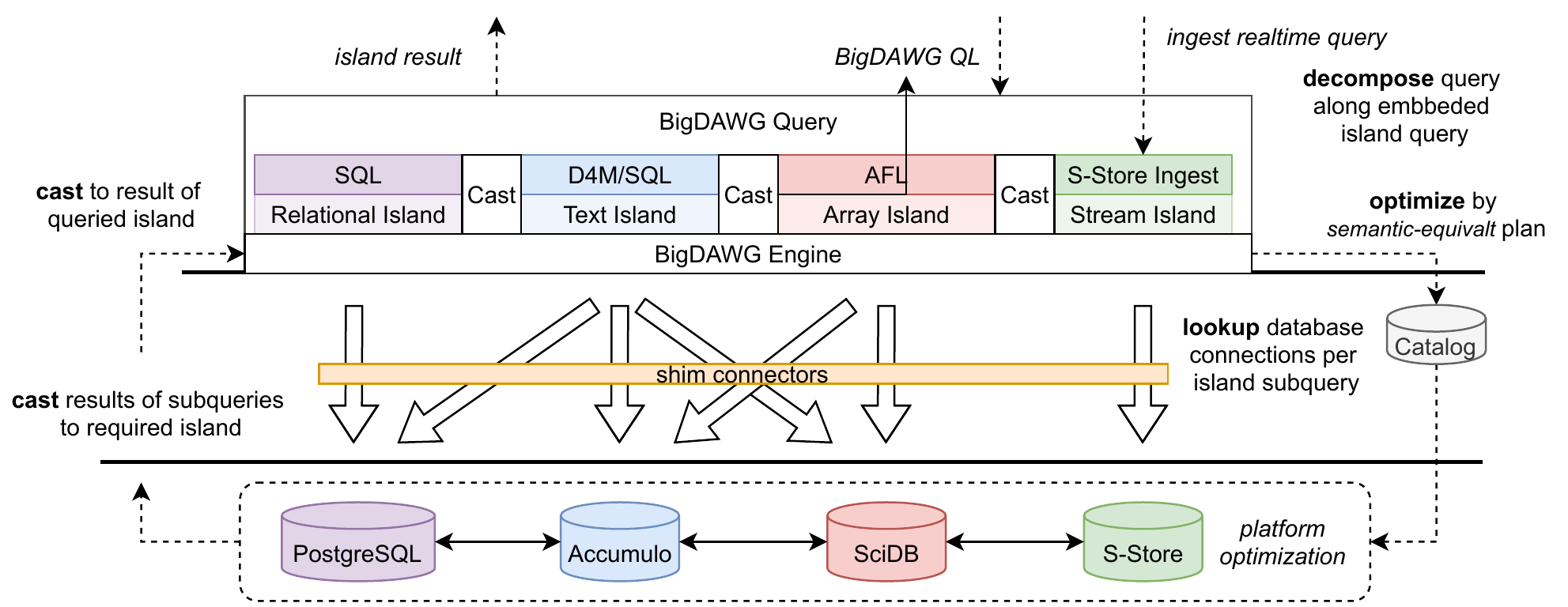}
    \caption{Conceptual architecture of the BigDAWG system, adopted from \cite{duggan2015bigdawg}}
    \label{fig:BigDawgConcept}
\end{figure}

\begin{figure}
    \centering
    \includegraphics[width=0.75\textwidth, trim=0 175 0 0, clip]{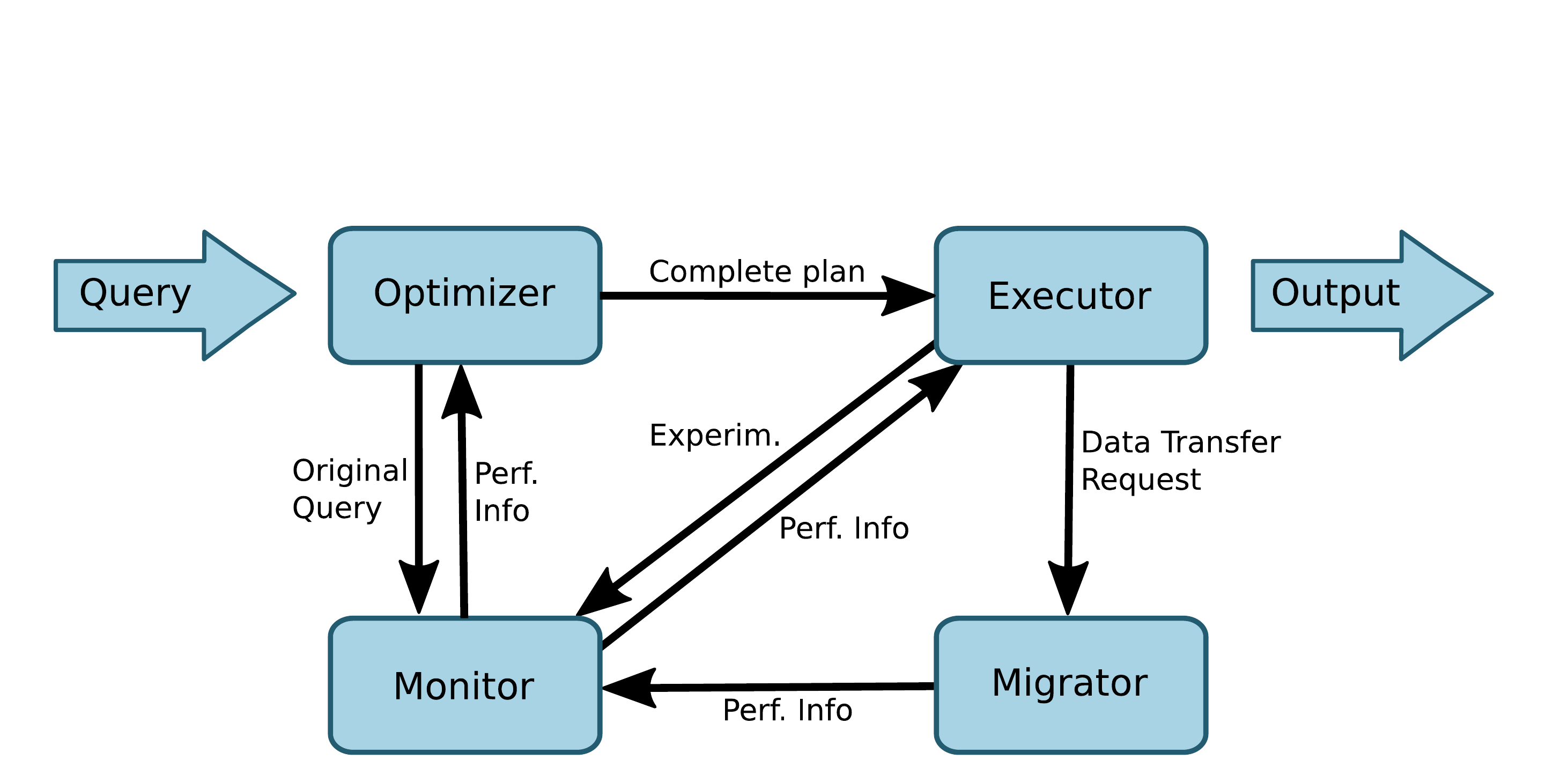}
    \caption{Technical architecture of the BigDAWG system, adopted from \cite{duggan2015bigdawg}}
    \label{fig:bigdawg}
\end{figure}

\emph{BigDAWG}\footnote{Big Data Analytics Working Group} is an extensible polystore system with support for multiple data models, real-time streaming analytics, and different visualisation interfaces \cite{duggan2015bigdawg, gadepally2016bigdawg}. The main goals of the system are: \begin{inparaenum}[(1)] \item the support of location transparency and prevent the user from managing an own mapping of data to store, \item providing complete semantics of each query concept, and \item enable the user to convert data from one query interface to another.\end{inparaenum} Figure~\ref{fig:BigDawgConcept} and Figure~\ref{fig:bigdawg} shows the technical design of BigDAWG, consisting of four distinct layers: database and storage engines, so-called \emph{islands of information}, middleware and API and the applications themselves.

Currently, PostgreSQL, Apache Accumulo, SciDB and S-Store are integrated into \emph{BigDAWG} and can be accessed using different \emph{islands of information}. An island consists of a data model, a query language and a set of storage engines. It enables the user to query a collection of these storage engines with a single query language. So-called \emph{shims} provide the mapping between the data model of the island and the underlying data models in different stores, allowing the extension to other databases by implementing the \emph{shim} contract. For this reason, \emph{BigDAWGs} support multiple islands other polystores such as \emph{Myria} (\ref{sys:myria}), \emph{D4M} as processing systems accessing Accumulo, SciDB and employing other semantics build onto of associative arrays. These islands do not procure the complete functionality of the underlying data stores. Therefore, in order to achieve semantic completeness, \emph{BigDAWG} initially comes with islands (\emph{degenerate islands} that provide the complete functionality of one data store. Additionally, it is possible to define new or existing islands by writing a shim for a new data store. Working with \emph{BigDAWG}, the user is also enabled to write queries accessing multiple islands by using so-called \emph{cast} and \emph{scope} operations.

The \emph{BigDAWG Layer} is a middleware responsible for receiving queries, query planning, query optimisation and benchmarking. Query processing works as follows: first, the query planning module parses the query, creates possible query plan trees, and assigns sub-queries to storage engines. The query plan trees are sent to the performance monitoring module, which determines the best tree based on information gained from benchmarking. At last, the query execution module uses the data migration module to identify the best method to combine the results and execute the query.

Query optimisation in \emph{BigDAWG} is based on a black-box approach and continuous monitoring of the performance of each query. The basic idea of optimisation is to move data only in case of expensive operations where a high-performance gain (at least an order of magnitude) can be expected. In order to minimise data movement, all local computations are performed first. Then, the results are combined. Apart from this basic optimisation, \emph{BigDAWG} collects information on the duration of (sub-)queries on various data stores. The resulting preference matrix is learned during three different modes: training mode, optimised mode and opportunistic mode. Every possible query plan is run in training mode on every possible storage engine. In optimised mode, \emph{BigDAWG} chooses the most promising execution plan for the (sub-)query based on the preference matrix. A random plan is selected if the matrix does not contain a similar query. The system also chooses an execution plan based on the preference matrix in opportunistic mode. Additionally, (sub-)queries might be evaluated further during times of low system utilisation or if new storage engines become available.

An additional optimisation aspect, \emph{BigDAWG} considers is the placement of data. Objects can be moved for load-balancing and optimisation purposes but only to data stores with \emph{shims} connected to all islands of source one. The movement of data is triggered by the comparison of the runtimes to similar queries over time, whose access all are going through the \emph{BigDAWG} middleware.

\emph{BigDAWG} has already been implemented and is available on GitHub\footnote{https://bigdawg.mit.edu/get-bigdawg}. Furthermore, the usefulness of the system has been evaluated for several use cases such as medical application\cite{elmore2015demonstration} an ocean metagenomic analysis \cite{mattson2017demonstrating}. In the medical application, we already discussed in Section \ref{sec:UseCaseMed}, \emph{BigDAWG} used PostgreSQL and Myria for clinical data, Apache Accumulo for text data, SciDB for historical waveform time-series data, and S-Store for streaming time-series data \cite{meehan2016}. A proposed polystore benchmark made evaluations in \cite{karimov2018polybench} showing the flexibility of the island concept and runtime improvements compared to Spark. Improvements reduce latency and benefits for migration tasks by moving data to an applicable store instead of using a reformulation of desired semantics in Spark.

\subsection{FORWARD}

\begin{figure}
    \centering
    \includegraphics[width=0.8\textwidth]{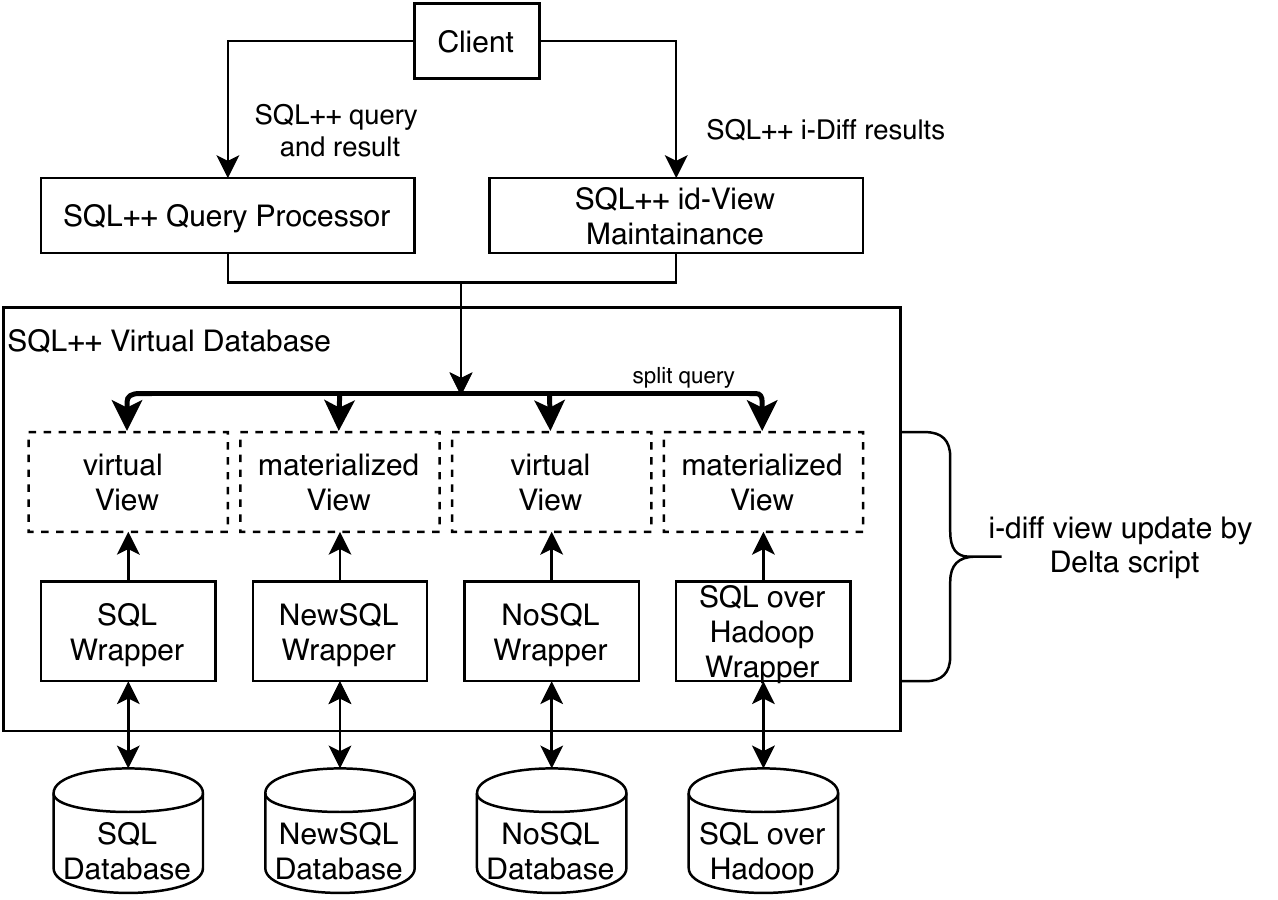}
    \caption{Placeholder for the FORWARD system according to \cite{ong2014sql++}}
    \label{fig:forward}
\end{figure}

FORWARD is a federated query processor that hybridises the different data models. The generic query language SQL++ \cite{ong2014sql++} implements a superset of SQL and consolidates different NoSQL query concepts within a semi-structured data model that integrates both JSON and the relational data model.

Tow major components compromise the middleware: the integrated view management provides virtual and materialised images. The SQL++ middleware uses the view concept to transfer the virtual view representation from the various stores into the data model of SQL++. Wrapper translates data queries on virtual views into the native query language. Materialised views preserve data states, besides virtual-only views, reduce the translation. Each update of these materialised views is employed by \emph{change} sets, for which FORWARD implements a different view-maintenance approach. The approach, called \emph{idIVM} for ID-based view updates, determines state changes between the underlying schema and the materialised view in FORWARD algebraically \cite{katsis2015utilizing}. Changed objects are formally defined by their key attributes from the original data sets. Conventional tuple-based comparisons are more compact and calculated more efficiently since not the entire tuple is compared, but solely things according to the key attribute. 

\emph{idIVM} considers four separate steps: within a $\delta$ script generation, it first determines the key attributes held in an i-diff schema for the original tables. Incoming changes are transferred to the i-diff values using entries from the modification log. This i-diff propagation determines the actual state change for which a series of DML operations are generated from the standard SQL in the script. Change operations such as DELETE/INSERT/UPDATE affect the materialised views and are optimised in a final step by performing a set of rewriting rules for each query. Semantic minimisation eliminates inefficiencies, composing individual operator rules to gain more performance.

As a hybrid approach to creating a uniform access layer for heterogeneous data, SQL++ considers variable semantics per query. Thus, various semantics can be defined as annotations to consider divergent processing of \emph{null} or equality operations. Throwing of an exception, the processing as zero or using a default value, e.g. with \emph{false}  for equality, are conventional solutions.

Internally, SQL++ queries are decomposed into individual sub-queries and use the maintained views. Partial results are integrated within the FORWARD Engine. A description of the translation into the native queries or the kind of decomposition is not provided.

SQL++ is currently implemented in \emph{AsterixDB} as well as \emph{AWESOME} and provides  SQL++ runtime system. It provides read requests and a schema definition language defined on SQL++, which allows defining individual types according to the JSON data model, in which a custom type-system implementation makes consolidation around different types.

\subsection{MuSQLE}
\emph{MuSQLE} is a distributed SQL query execution, targeting multi-engine environments \cite{giannakouris2016musqle}. 

MuSQLE is a SQL engine to execute relational queries on multiple platforms, able to move data between them. It allows the reformulation of SQL queries and assignment of subqueries on a fixed set of supported engines. In doing so, MuSQLE pays attention to reducing the overhead of data movement by keeping operations on data as much as possible together and selecting a join order afterwards according to the locality heuristic.
The query engine currently supports PostgreSQL, SparkSQL and MemSQL, all with native support for SQL dialects. It implements partial aspects with schema administration, query planning and execution, strictly based on the relational model. The support by MuSQLE is the distribution of query execution and federation according to \ref{sec:System types} over this heterogeneous landscape. MuSQLE optimises the SQL execution by intercepting the Spark interfaces via a proposed API contract for extending the SQL distribution on other systems. 

System optimisation is performed on the logical plans, allowing local physical optimisation by engines and taking intermediate result movements into account. Although each engine operator and cost model do not need to be integrated, only specific API calls must be implemented. The changes made to engines allow the creation of virtual tables and estimate execution time for queries.

MuSQLE consists of three central components. The \emph{Metastore} component stores and manages the schema and mapping information for each table. Tables are stored distinctively in one of the supported SQL engines as a whole. The \emph{SQL Parser} reads and validates the query and generates the corresponding object model in the form of a query graph. The multi-engine optimiser uses a cost model defined explicitly for MuSQLE, and the \emph{Engine API} acts as a wrapper layer in which the communication with the engine is abstracted, or the query is delegated further.

Separated into two categories for execution and estimation, the API for MuSQLE offers the possibility of calling up the actual execution of a given SQL query and time estimation for each connected engine. The query planning uses this engine API, coupled with the concrete SQL engines and expects five functions that must be implemented individually in each wrapper. In the current implementation, MuSQLE checks queries in PostgreSQL and MemSQL by utilising their available EXPLAIN statement, while for SparkSQL, multiple individual cost models are implemented for each SQL operator.

In addition to cost estimation for query execution, estimates for migration within the system are also determined. The emph{Engine API} provides a function in which the loading of tables, encapsulated in MuSQLE by an engine-spanning Spark DataFrame, can be determined.

MuSQLEs optimisation focuses on the JOIN order and uses the \emph{DBSize} dynamic programming planning algorithm already known from DB2 with its extension as \emph{DPhyp} in \cite{nica2011call}.

\emph{DBhyp} generates a linear JOIN graph (left-deep-join tree) which is iteratively enlarged. The choice for the next JOIN partner is examined by counting the linked subgraphs. Such a connection exists if, for two induced sub-graphs of the JOIN graph, each with a disjoint set of nodes, at least one edge exists which connects the two sub-graphs. These are called connected sub-graph pairs (\emph{csg-cmp-pairs}). MuSQLE extends this JOIN order planning by a location-based optimisation and the inclusion of cost and statistics estimates. 
In location-based optimisation, the properties of the distributed tables and the transfer of intermediate results are taken into account. This is achieved by determining for the DBhyp algorithm a plan for precisely one existing engine and all available ones.

The enhancement achieves the inclusion of costs and statistics using the available information from the Engine API in DBhyp. Therefore each possible execution plan is checked with its engine combination and required data movement operation, utilising the table loading cost function provided by the API. Estimated costs are determined for each join using the identified csg-cmp-pairs of the extended DBhyp.
Besides, MuSQLEs query planning checks to shift the query processing to another SQL engine that is cheaper execution. Considering the costs for loading intermediate tables into the engine and preparing temporary tables, MuSQLE checks all possible execution plans.

Since MuSQLE provides individual cost estimations of relational operations to integrate SparkSQL, new systems must implement the MuSQLE API engine contract. This contract requires the cost estimation for relational processes and the interpretation of the SQL language.

Additional effort in query planning is required due to the extra search dimension coming through each engine owned performance for a specific query.

Experimental evaluations with MuSQLE successfully transfer whole intermediate table results and benefits in location-oriented planning for the TPC-H benchmark. However, it is prohibitive for using it with large datasets required in big-data analytics.

\subsection{RHEEM}\label{subsec:RHEEM}

\begin{figure}
    \centering
    \includegraphics[width=\textwidth]{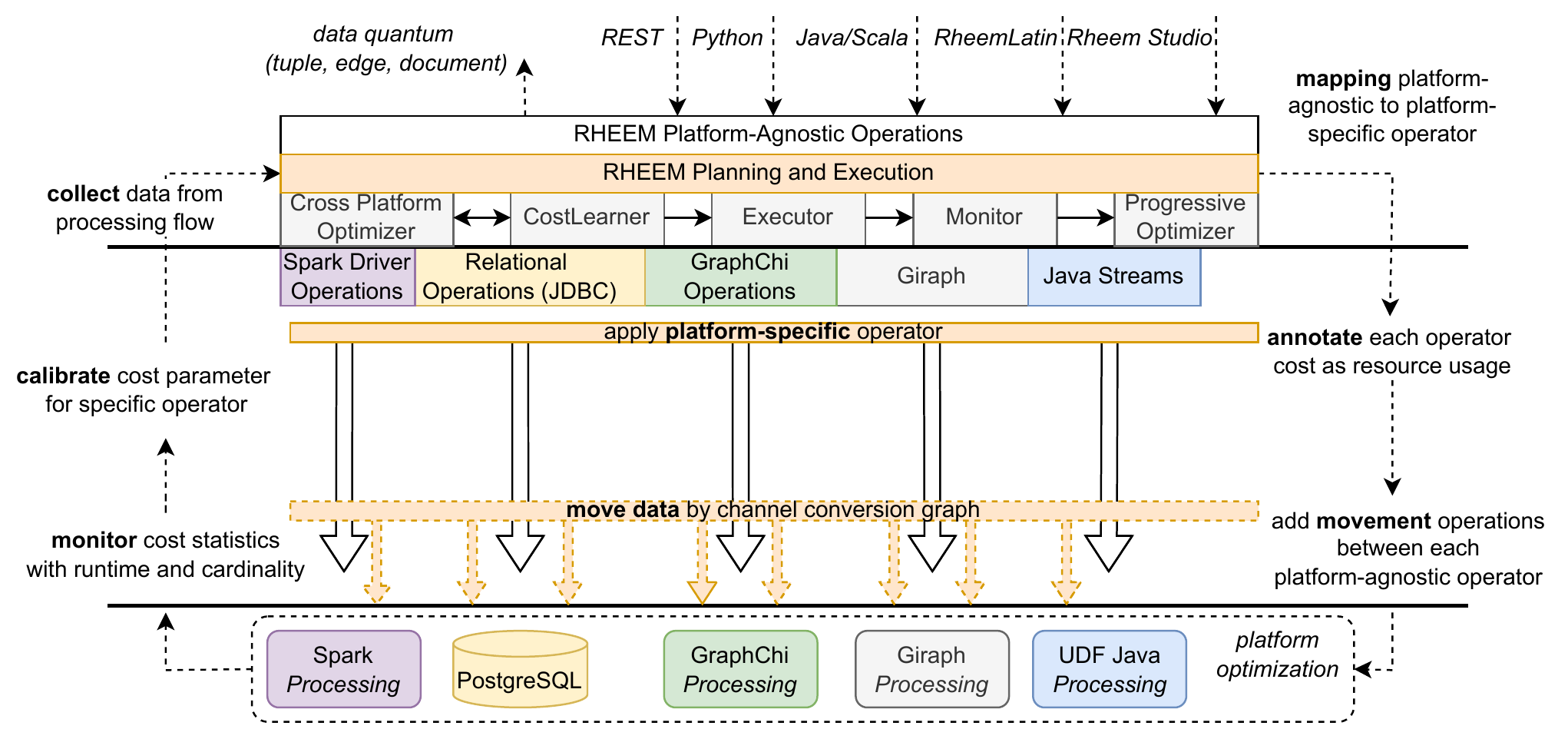}
    \caption{Technical architecture of the RHEEM processing system, adopted from \cite{agrawal2016rheem} with data storage and processing systems.}
    \label{fig:rheem}
\end{figure}

\emph{RHEEM} is general-purpose cross-platform data processing system (DBMS, MR, NoSQL) objecting to decouple the application by providing a multi-platform execution and data storage independence \cite{agrawal2016rheem}. The goal of \emph{RHEEM} is efficient utilisation of so-called tasks in connection with the migration of dependent workload within a distributed system. Due to the high variance of data models and database systems through their API, \emph{RHEEM} propagates a 2-tier model. On the upper layer, \emph{RHEEM} uses platform-agnostic operators for mediation and migration operations, whose accessing application makes use of them and provides an abstraction to the data. On the lower layer, \emph{RHEEM} uses platform-specific operators that are executed directly on the dynamic data processing systems. For this purpose, the system maps platform-agnostic operators to platform-specific execution operators, including analytical operators such as map, reduce, group-by, and reformulated equality and inequality joins. \emph{RHEEM} uses specific platform implementations and connects them through the internally used Java runtime platform. The mappings can be either direct (1:1) to an underlying platform-specific operator, extending (1:n) to several operators or reducing (n:1) if one operator already fulfils all functions in combination. 
The set of possible combinations of input and output operators is the set of all available executions, in which \emph{RHEEM} calls an inflated plan. This inflated plan contains connected operations and the partitioning of required data.

In contrast to other systems such as ESTOCADA \cite{alotaibi2019towards} or FORWARD \cite{ong2014sql++}, \emph{RHEEM} calculates no materialised views and only keeps the data where it has initially been stored, moving data objects temporarily for the scope of a query. In particular, the migration between platforms in \emph{RHEEM} is possible by utilising a so-called \emph{conversion graph}, whose vertices contain the different data formats/structures and directed edges consider the transformation from one model to another \cite{kruse2019optimizing}. Self-defined or existing operations implement the so-called conversion channels as model transformations steps and are reusable (e.g. service or mapping functions) or not (e.g. streams when they are at the end). The direction of the edges describes clear possibilities to translate from one model to another. The goal of \emph{RHEEM} migration is to create a minimum conversion tree (MCT) from the possible conversions in which the root format leads to one or multiple consumers. The concrete problem is optimising and reducing the conversion cost, considering the reusability through the directed channels. In \cite{kruse2019optimizing}, it was proven that the problem in finding the MCT is NP-hard. 

To produce an efficient execution plan from the inflated plan, \emph{RHEEM} estimates the cost for each platform-agnostic operation and cardinality with necessary migration steps. The system annotates the inflated plan with these estimates, extended by the costs of necessary migrations, which arise from the MCT and concretely determines the resource utilisation from a genetic cost model. This cost model considers a platform-agnostic operation with estimated cardinality and calibrates parameters for different acquired resources. By monitoring and collecting execution statistics of a given plan, \emph{RHEEM} differs the estimated cost from the actual one, calculating a geometric loss for an agnostic operation. The loss improves the prediction by calibrating weighting factors for multiple used resource features such as CPU cycles, consumed storage and outer resource utilisation (memory, disk, network). Calibrating these low resource cost factors requires an abstraction of processing to a single kind of data object. For this reason, \emph{RHEEM} introduce an additional abstraction level, given for concrete data objects, so-called \emph{data quantum}. Data quantum can represent multiple formats such as database tuples, graph edges, and document content, helping to trace the resource consumption when processing individual objects.

When extending \emph{RHEEM}, newly supported systems must provide a mapping of the abstracted \emph{RHEEM} operators to their local operations or operation sequences. These types must be integrated into the conversion graph if new systems support new models, respecting the input conversion channels.

\emph{RHEEM} provides a development environment \emph{RHEEM Studio} \cite{lucas2018rheemstudio} for non-experts in order to create \emph{RHEEM} plans, using their processing language \emph{RHEEMLatin} and other interfaces such as Java, Python or simple REST. According to  \citet{agrawal2016rheem}, the ecosystem supports Postgres, JavaStreams, Apache Flink, Apache Spark, GraphChi and Giraph.

\section{Related Systems}\label{sec:RelatedSystems}

In our research on polyglot persistence, we encountered multiple systems which fall into or near the category of polyglot managed systems but did not fit our criteria of a data store very well. These systems focus more on other aspects of data management that we do not consider in the scope of this paper (such as pure OLAP systems, query engines or frameworks). In this section, we want to give them some spotlight and briefly present their idea and purpose but will not go into detail for our analysis and comparison further on (see Section~\ref{sec:Comparison}).

\subsection{Apache Calcite}

Apache Calcite \cite{begoli2018apache} is a Java-based software framework that provides query processing, optimization, and support for the SQL query interfaces, granting uniform access to multiple data processing systems. 

Calcite's architecture consists of a modular and extensible query optimizer with built-in optimization rules. The query processor handles a variety of query languages, utilized by an adapter architecture designed for extensibility and support for heterogeneous data models. This adapter approach allows the user to connect data stores containing a subset of data objects. Internally, Calcite represents its queries via a relational operator tree, implementing the base SQL language with a subset of extensions. This subset concerns stream-based queries such as sliding windowing expressions and geospatial queries. A dedicated GEOMETRY data type enables geospatial queries. Stream-based queries are implemented internally by a dedicated window operator, and a new map and array type allow to access semi-structured data.

\begin{figure}
    \centering
    \includegraphics[scale=0.3]{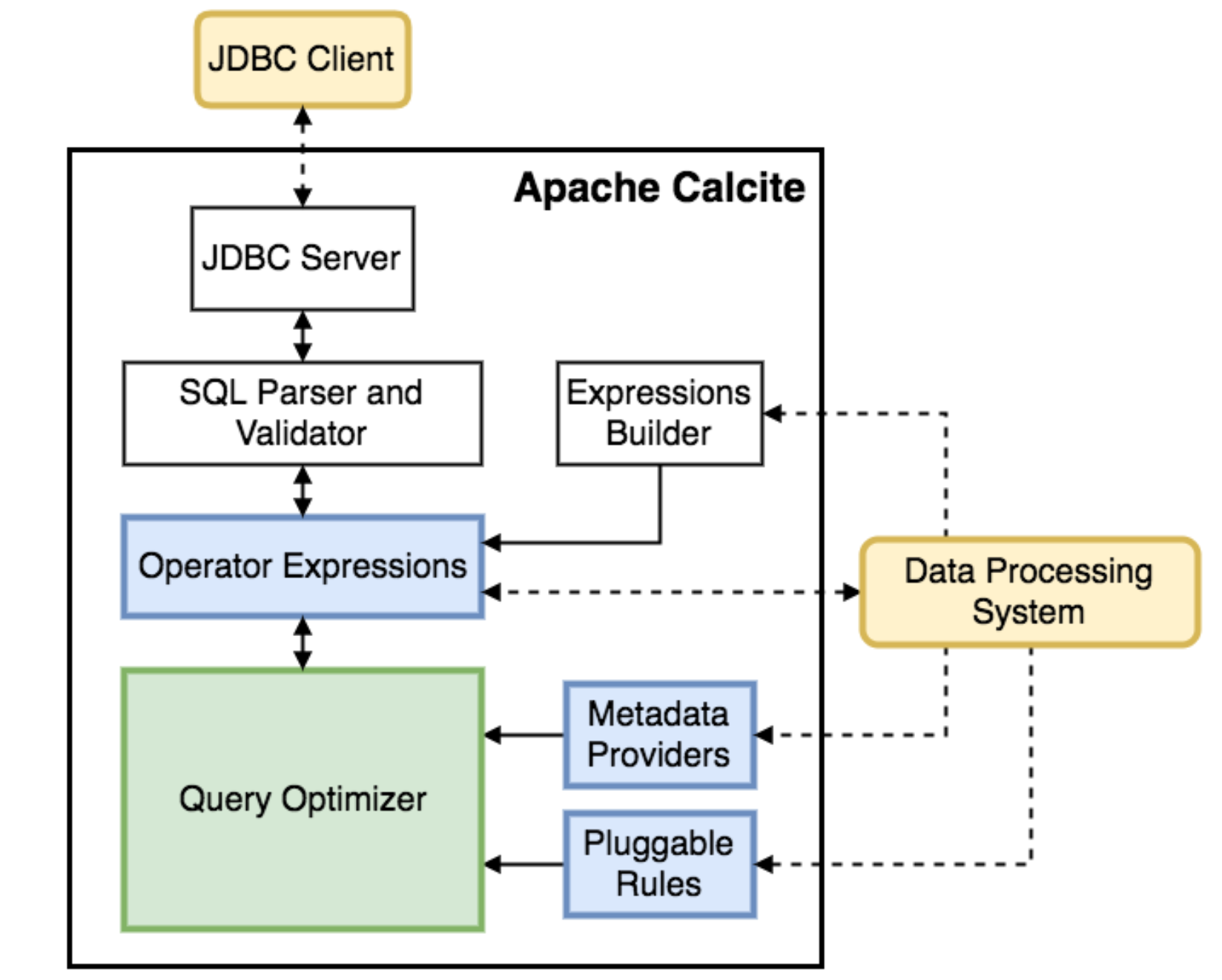}
    \caption{System integration architecture for query planning and optimization (based on \cite{begoli2018apache})}
    \label{fig:apachecalcite_architecture}
\end{figure}

Apache Calcite is available as a framework, and the framework design is required to define its mediation system with configuration concerning connected stores with Local-as-View mapping of source to the global one. Figure~\ref{fig:apachecalcite_architecture} shows the integration architecture of Calcite with query processing components and required user interaction to couple external data systems.

The central component of Apache Calcites is its query optimizer, consisting of planning rules, metadata providers and a planning component. The set of planning rules includes transformations of the query tree, in which possible transformations preserve the semantics of the query and resolve via pattern matching on the operator tree. Calcite includes multiple optimization rules for respective connected sources, containing precise query reformulations whose set can be extended manually for each connected system.

The interaction between the framework and a system is possible in Calcite in several ways. When using Calcite for heterogeneous access, an available common compliant JDBC interface allows expressing the SQL statements directly, whereas the support for LINQ4J language extension in Java abstracts completely from SQL queries. LINQ4J encapsulates the query formulation behind Java functions and delegates, letting the user work within a single programming environment. 

In addition, Calcite allows to link  Or the framework can be linked over so-called Calcite adapters with new systems, in which a closed converter sub-component is implemented as an interface. This converter is responsible for the transformation into the native interface and the forwarding of the relational expression.

The framework is under active development and is currently being used in multiple public projects such as Apache Beam, Flink, Drill, Solr, Phoenix, Apex, Storm and Samza.

The framework omits some key components, e.g., storage of data, algorithms to process data, and a repository for storing metadata. These omissions are deliberate: it makes Calcite an excellent choice for mediating between applications having one or more data storage locations and using multiple data processing engines. It is also a solid foundation for building bespoke data processing systems. The Calcite architecture is not only tailored towards optimizing SQL queries. In contrast, data processing systems commonly use their parser for their query language.

\subsection{Apache Drill}

Apache Drill is part of the Apache Software foundation's software project portfolio and aims at providing an open-source schema-free SQL query engine for Hadoop, NoSQL data stores and cloud storage solutions \cite{onlineApacheDrill2020}. As stated in their initial proposal for the project, the team behind Apache Drill took inspiration from Google Dremel to provide an open-source solution capable of scaling to 10.000 or more servers and processing petabytes of data \footnote{\url{https://cwiki.apache.org/confluence/display/incubator/DrillProposal} (last accessed: 06.12.2020)}. Thus, Drill supports a variety of NoSQL databases and file systems, including HBase, MongoDB, MapR-DB, HDFS, MapR-FS, Amazon S3, Azure Blob Storage, and Google Cloud Storage, Swift, NAS, and local files \cite{onlineApacheDrill2020}.

\begin{figure}
    \centering
    \includegraphics[scale=0.75]{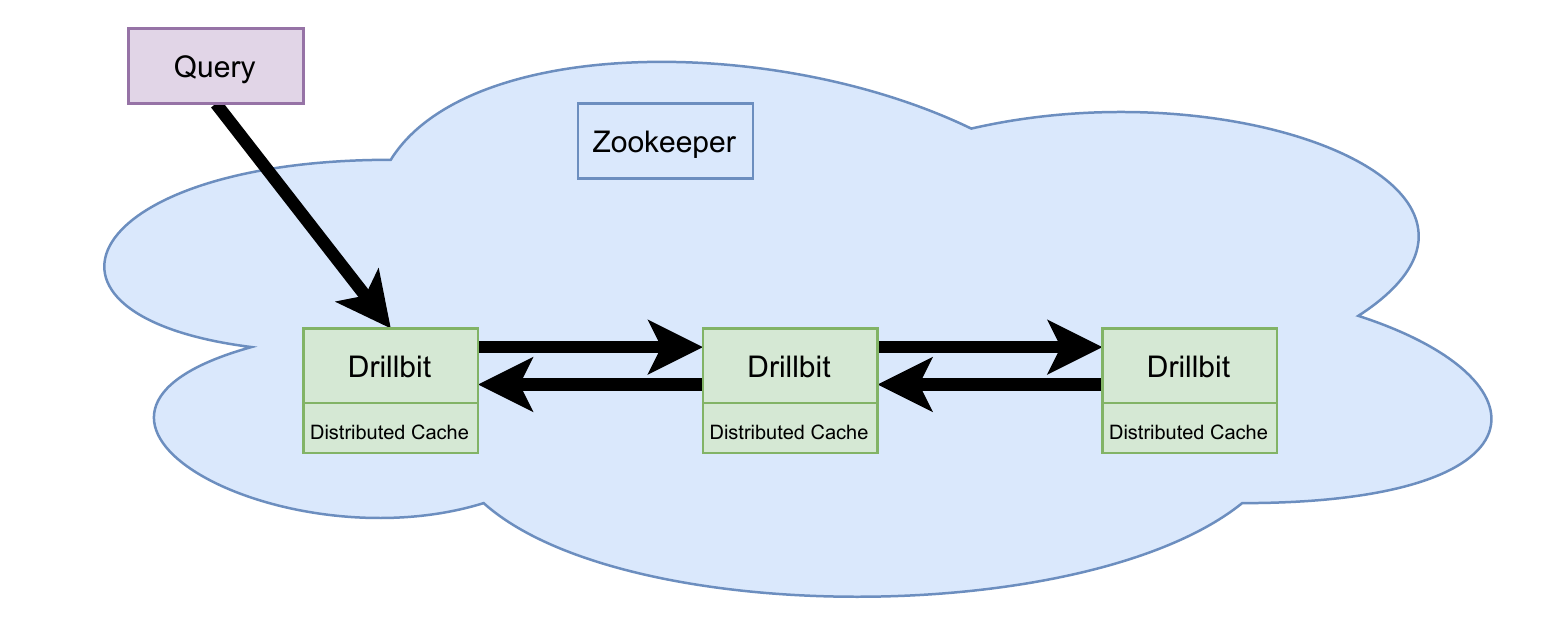}
    \caption{High level representation of the flow of a Drill query and the system's architecture (based on \cite{onlineApacheDrill2020})}
    \label{fig:apachedrill_architecture}
\end{figure}

The main component of Drill is the Drillbit service. This service is part of Apache Drill's concept of a distributed processing engine using multi-level serving trees \cite[p.~58--59]{mazumder2016}. Every node running a Drillbit service is capable of
\begin{inparaenum}
    \item accepting a query and \item performing a query rewriting process.\item Generating an optimized execution plan and 
    \item run the plan on appropriate nodes\footnote{A Drillbit service consults a Zookeeper instance to check for suitable and available Drillbit nodes},
    \item collecting the results of all involved nodes and finally
    \item sending the client the queried result.
\end{inparaenum}
Figure~\ref{fig:apachedrill_architecture} illustrates the flow of a Drill query and the system's architecture on a high level.

Apache Drill can join data from multiple datastores in a single query without the client's need to manage metadata. This information is automatically derived from Drill's storage plugins to integrate the different datastore into its system. These plugins also enable Drill's datastore-aware optimizer, which is necessary for the query optimization process of a Drillbit service, in which Apache Drill considers data-locality for their Drillbit service, colocating the service on the same node where the datastore is running.

\subsection{AWESOME}

\emph{AWESOME} (\emph{Analytical Workbench for Exploration of Social MEDia}) \cite{dasgupta2016analytics, dasgupta2017generating} is a multistore system focusing on data investigation and analytics. Currently, \emph{AWESOME} supports relational, semi-structured and property graph data as well as a set of computational structures for matrices and vectors. The data stores which are integrated into the \emph{AWESOME} system at present are PostgreSQL, AsterixDB, Neo4j, Apache Solr Cloud and a Spark computation engine. Replacing these data stores is possible if the new data store has the same model. Then, a couple of query translation rules must be redefined according to the new data store.

Together, the different data models and data stores allow \emph{AWESOME} to handle two kinds of data heterogeneity:
\begin{inparaenum}
    \item applications that have to store and query heterogeneous data sets and analyze them by combining different data models in arbitrarily complex ways and
    \item applications where data sets have a mixed model.
\end{inparaenum}

Data placement and computational processes of raw and derived data are defined by the data ingestion language ADIL. This \emph{AWESOME} specific language allows specifying complex ingestion workflows, ingestion filters and complex query statements. For instance, it enables the user to define store specific data acquisition processes and route data into appropriate storage. Furthermore, the construction of \emph{analysis environments} and data augmentation processes is possible. The developers of the \emph{AWESOME} system differentiate between \emph{dictionary-based} and \emph{computational} data augmentation. In the case of dictionary-based data augmentation, a polymorphic join operation is performed periodically between terms in a dictionary and any data set residing in one of the underlying data stores. On the other hand, computational data augmentation is achieved by passing a data item through a function that returns the augmented for of the item as a result.

The architecture of the \emph{AWESOME} system consists of nine different components. Unfortunately, the authors do not provide an architectural overview of the system's components or further information on their query processing workflow. Nevertheless, some components are mentioned and will be described below.

One of the main components is the \emph{Data Ingestion Module}. It receives data from various sources (e.g. file sources, database sources, data APIs and streaming APIs). The connection between the system and the different data sources is realized by adapters that can gather and filter data and perform manipulations. Furthermore, the adapters are responsible for populating data based on a decision table into the appropriate store. The \emph{Data Derivation View Manager (DDVM)} uses database views and model transformations to manage the data derivation process. It can be considered as a repository for derived views that compiles view definitions into workflows of tasks. Besides, the \emph{DDVM} ensures that the materialization processes of every view continue to be executed if updates (insert-only) occur in the primary repository. The \emph{Query Processing Module (QPM)} parses, plans, optimizes and executes queries. These queries may either be pass-through queries that are expressed in the native query language of an integrated data store or \emph{AWESOME} queries that are processed through a data federation layer using a hybrid data model. The \emph{Data Computation Module (DCM)} is responsible for the computational processes and manages the lifespan and availability of results derived from complex ingestion workflows. In case of user-defined functions that cannot be computed inside any store, the \emph{DCM} manages necessary data exchanges and returns or stores results. 

Apart from these main components, the \emph{AWESOME} system consists of an internal data store that is used for objects that cannot be stored in the integrated data stores, an additional cache for data that can only be stored and retrieved but not queried and a \emph{System Catalog (SC)} that keeps track of all data sources and derived views. The integrated \emph{Performance Monitoring Module (PMM)} monitors the system resources such as memory, number of threads and IO operations so that the ingestion and query processing modules are able to incorporate this information. Furthermore, third-party analytical libraries provide access to text processing, graph, statistical, and machine learning methods.

Currently, complex ingestion processes and queries have to consider the native data models of the integrated stores, the different query and ingestion capabilities, and the different computational costs for data insertion, fragmentation, filtering, index creation, data movement, and query processing and computations. Here, the authors see room for improvement so that, in their future work, they plan to enhance the inflow optimization, the content optimization and the dataflow optimization. Inflow optimization decides where to store the data and how often they need to be replicated. In contrast, content optimization covers the elimination of content-free, insignificant or redundant data, cross-store information deduplication, integrity constraints and clustering. The authors plan to use a data flow graph to optimize the data flow and adjust their computed plans to the overall system's load (burst and quiet periods).

\subsection{DBMS+ (Cyclops)}

DBMS+ (Data Management System for Multiple Systems)\cite{lim2013fit} is a conceptual approach proposed in conjunction with the existing Cyclops platform to enable distributed or centralized query processing on polyglot system landscapes. The approach provides the user of a domain to specify its application, including availability, consistency, changeability, and cost requirements for the DBMS+. In request processing, these restrictions are taken into account, generated, and evaluated as so-called multi-system execution plans by the DBMS+. In terms of content, the thesis illustrates its approach by employing continuous data stream queries on several analytical example scenarios related to interactions between users and web pages. 
All scenarios are covered in the concept by continuous data stream queries. In its concept of continuous stream-based execution, DBMS+ first distinguishes between centralized processing, where exactly one processing system is used, and distributed processing, where several different systems are used. In terms of execution types, the plans distinguish between \emph{incremental} and \emph{non-incremental} execution plans. Non-incremental execution plans extract data objects from a request window on the stream and process them directly via a group-by aggregation. On the other hand, incremental processing caches aggregations that have already been executed and pass them on to the next operation only when the result is complete. Furthermore, DBMS+ distinguishes between \emph{imperial} and \emph{federated} processing in a distributed execution. \emph{federated} processing uses the native query language of each connected system, leaving independent parsing and local optimization to the individual databases. This independence makes it easier to connect the systems to the higher-level mediation and exploit local, potentially costly optimizations for each system. However, semantic differences due to different languages must be reconciled.
On the other hand, the \emph{imperial} processing called in DBMS+ avoids considering a native query language and skips parsing, generating local plans, and optimizing. It interacts directly with the execution and storage component of the respective system (for example, in the case of the public-access storage component in MySQL), giving it complete control over all storage operations. This direct access avoids constraints on the respective systems and semantic inequalities.
Open questions consider implementing and integrating the proposed DBMS+ approach into the existing Cyclops system. Especially which plans should be identified and selected for these connected systems and how the above requirements must be structured and described to get an interpretable format. Furthermore, problems from distributed data management have been discussed that concern the provision of storage space and how related data can be stored in a distributed manner.

\subsection{Odyssey}
Odyssey \cite{hacigumucs2013odyssey} system is a multistore for handling multiple analytical processing systems for big-data queries in Hadoop and parallel OLAP. It enables the storage and retrieval of data within HDFS and RDBMS using opportunistic materialized views based on MISO \cite{lefevre2014miso}. MISO is a method of tuning the physical design of a multistore system (Hive/HDFS and RDBMS) by deciding to place data in multiple data stores, thereby minimizing necessary data shifts between intermediate result stores. The results of query execution are treated as opportunistic, materialized views that can then be located in the underlying stores to optimize the evaluation of subsequent queries. Thereby MISO builds a sort of "caching" for partial results of outbound analytical calculations. MISO aims to maximize the performance of ad-hoc processing of large data queries by deciding where data is best stored. Therefore, it must maintain and calibrate its cost functions for estimating the cost of operations. As a result, there is an integration effort to generate the corresponding cost estimates for each new engine added.

This optimization problem is equivalent to the Knapsack problem, which is intensified by two dimensions. Each Knapsack has two-dimensional capacities with current storage capacity and transfer quotas. In addition to considering the current workload, future transfer costs must also be taken into account through subsequent queries, which are optionally refined by user-defined constraints. For example, some data can be persistent only to a particular format without changing the semantics. Specifically, MISO refers to this problem as a Multistore Design Problem (MDP), where an observed stream of queries on the current multistore design is used to determine a new one that meets the specific constraints and reduces workload costs.
In practice, a so-called MISO Tuner component performs a periodically recurring (online) optimization of the current workload distribution.

\subsection{QUEPA}

QUEPA is a tool for data exploration and augmentation within polystore setting \cite{maccioni2016quepa}. It shows a possible contribution to this problem by introducing augmented search and augmented exploration, two new methods to access a polystore based on the automatic enrichment of data extracted from a database with data that belong to the rest of the polystore. Data augmentation search considers the stepwise expansion of search results over a local database with relevant data for this query stored elsewhere in the polystore. The exploration exploits the same infrastructure and provides an interactive way to access data, similar to searching the Web. 

QUEPA proposes the following three components: A repository, implemented as a Neo4j database, stores relationships among related data objects in the polystore. The repository is an undirected graph in which the vertices represent the same queried conceptual model. References among these objects are represented as edges in the graph equivalently. Each database serves one or many collections (e.g. tables in a relational database)containing the undirected references between those objects, restricting the data types only to those systems whose items are identifiable by a name and a key.

The collector must identify the relationships between data objects stored in the repository. Due to inherent heterogeneity given by the polystore with their partially consisting schema-less databases, QUEPA combines various unspecified entity resolution approaches from literature in an offline manner. For schema-bound data and information, the system follows proven techniques. Therefore QUEPA considers an alignment- and a record-linkage phase (for resolving entities and removing duplicates) on an existing schema. 

The third component is responsible for request processing based on data augmentation, in which incoming requests are validated and rewritten. QUEPA first checks the type of query in which aggregate-oriented queries are executed entirely without augmentation. The query processing adds new data objects based on the identified links with the same reference level from the repository. This query method makes it able to traverse the data and retrieve them in an exploratory way.

This feature distinguishes QUEPA from the middleware approach in that no additional abstraction layer is created using a mediator. It uses a gentle method for data integration and stores data objects in native format.

QUEPA currently exists as a prototype implementation with multiple stores covering the most used data models by MongoDB, MySQL, Neo4J and Redis database. The paper demonstrates the QUEPA system functionally with four different scenarios:\begin{inparaenum} \item simplicity of configuration and startup of QUEPA, \item query result comparison with and without augmentation, \item query result comparison with and without augmentation, \item a simple exploratory query and \item adding new system types and promotion of new relationships.\end{inparaenum} The evaluation does not cover non-functional aspects such as consumed resources and query performance.

\subsection{QoX}

QoX \cite{simitsis2012optimizing} is a particular type of workflow-driven analytical engine that integrates data from relational databases, Extract-Transform-Load (ETL) tools and varying execution engines (such as MapReduce). QoX specifies data flow in an ETL-like workflow, allowing it to combine structured and unstructured data. It provides generic data flow operations such as join, aggregation and filtering, user-defined functions, and pre-defined analysis procedures (e.g. sentiment analysis or product identification). The system's primary focus is to optimize these integrating workflows in performance, freshness, latency, cost, and scalability. To enable the user to define optimization goals for their workflows, \emph{QoX} provides the feature to annotate the corresponding queries.

\subsection{TATOOINE}
\emph{TATOOINE} is a data integration tool that has been developed in collaboration with french journalists~\cite{bonaque2016mixed}. The resulting tool can combine several different sources and enables the user to analyse web and social network communication.

\emph{TATOOINE} uses a GAV data integration approach to store static as well as dynamic information from structured, semi-structured and unstructured databases or social feeds. Apache Solr, a set of relational databases and RDF data sources, is currently integrated. Since \emph{TATOOINE} does not use mappings or an integrated global schema, queries are evaluated over so-called \emph{mixed data instances}. \emph{Mixed data instances} consist of a set of data sources and an application-dependent RDF graph that contains domain-specific knowledge about the data within each store and the connections between data residing in different stores. Information and structures from the RDF graph can be combined with any data source using the join opportunities provided by repeated values such as names or hashtags. The resulting \emph{mixed queries} combine sub-queries expressed in the data store's native language with RDF querying. RDF query languages such as SPARQL provide, amongst others, operations like disjunctions, pattern matching, aggregations and the construction of triple results, as well as the derivation of implicit triples of the RDF graph.

Even though \emph{mixed queries} are a powerful and flexible tool, a vast amount of knowledge and expertise is necessary to use them. Therefore, \emph{TATOOINE} also provides keyword-based querying. For each source, digests are derived, containing the schema and a representation of a set of atomic values associated with positions in the schema. The digests can be viewed as directed graphs. Given two or more keywords, the engine looks them up in the digests and identifies the shortest paths between them afterwards. The shortest paths contain additional information that is closely related to the keywords.

\emph{TATOOINE} is strongly tailored to the specific usage patterns of journalistic research. In favour of more general approaches, it has not been analysed further.

\section{Comparison}\label{sec:Comparison}

Based on their different design objectives, the systems presented in Section~\ref{sec:Analyzed Systems} offer different properties, advantages and limitations. In this section, we categorize them according to the system types we presented in Section~\ref{sec:System types} if applicable and propose a comparison based on the concluded requirements we identified in Section~\ref{sec:Concluded Requirements}. The results of our categorization and comparison efforts are displayed in
Table~\ref{tab:Architecture}, Table~\ref{tab:FuncFeatures} and Table~\ref{tab:QueryProcessing}, featuring information about system types, the systems' architectures, functional features and query processing capabilities as well as their adaptivity properties.

\input{./tables/table_commands.tex}

\input{./tables/table_architecture.tex}

\input{./tables/table_functional_features.tex}

\input{./tables/table_query_processing.tex}

\subsection{Architecture}

The comparison in Table~\ref{tab:Architecture} shows significantly differences in system design for all considered systems, themselves stated as polyglot or multistore database or data management. Despite the wide acceptance of applying polyglot design in supporting use-case by a mix of database systems, there is still no common sense in designing such system kind overall. Only the classification by \cite{rahm2015verteiltes} in separated designs of linked-data, mediator-wrapper- or simply data warehouse-systems were being approved. Systems suited more public access result in linking multiple heterogeneous system by given a Local-as-View mappings and common ontology with unified access via a set of \emph{classes} and \emph{relationships}. Data access is applied by reasoning on the ontology, expressed with Description Logic what is called \emph{Ontology-Based Data Access}. Building polyglot systems from scratch results in concerning a distinct mediator using one ore multiple wrappers in order to query data from underlying data stores. In contrast to the linked-data approach, in which (web-)data bases on ontology's, stored in RDF format, queries are formulated against a single query interface or in this native language where data is located. A formulated query is either being executed directly without concerning any capabilities or planned and optimized, minimizing essential migrations steps. Given the single intermediate model, results are represented in this way how store communication is managed (e.g. exchanging complete relations or single tuple data objects).
Besides abstract design concepts, Table~\ref{tab:Architecture} shows that the common sense in distinctive concerning these described system types in Section~\ref{sec:System types} is mostly approved. The majority of systems can be categorized as multistore. Besides the simplicity of managing a single query interface, this approach allows to include capabilities of the connected databases in one's own query planning. In contrast, prominent references of BigDAWG~\cite{duggan2015bigdawg} and Polypheny-DB~\cite{vogt2018polypheny} do not aim at merging the respective capabilities of the databases into a higher-level model, but rather at advancing the respective available language. While BigDAWG does not include any data location in the planning and this must be managed and formulated by the user, Polypheny-DB has the overall goal of being able to apply all available query languages to the entire set of all managed data objects. 
A special focus is represented by the commercial solution from Microsoft in the form of PolyBase \cite{dewitt2013split}, linking, external data sources such AS HDFS cluster provider and CosmosDB with its own SQL server T-SLQ language and operation. Despite its simplified use case as a wrapper for reformulating SQL queries and evaluating them on mappings into external systems, this system is classified as polyglot.

The first evaluated dimension captures functional features. Polypheny-DB and AWESOME are the only data stores which support write operations, whereas all the other stores only provide read functionality and have their focus only in an optimized and integrated multi-engine analytics, which does not operate on a global schema. Only the AWESOME system is using their \emph{ADIL} language as a form of explicitly defining a schema. Furthermore, the lack of a DDL in most of the systems results in the non-existence of data constraints, which corresponds to the read-only character. Advanced features such as data privacy, security, as well as push-based queries, are provided by none of the systems.

FORWARD and Polypheny-DB support the most extensive amounts of varying data models, in which the latter is planning to enable the user to specify data requirements (e.g. a minimum level of consistency) in forms of annotations.
In terms of query processing, the adopted approaches vary strongly. One part of almost all query execution systems (except for CloudMdsQL and ESTOCADA) is query splitting. The optimization of queries is in the majority of cases cost-based, but heuristics and benchmarking are also applied. A common data model is used by all stores except for BigDAWG. Most of the systems use the relational model but, especially when NewSQL-stores are enclosed, associative arrays and documents are also common.
An aspect which seems to be slightly neglected is the semantic of the systems or rather the semantic mapping between different data stores. For almost all cases, the semantic mapping is fixed or not even considered at all (also referred to as fixed. FORWARD offers with its SQL++ query language the most complex approach where the semantic mapping between the underlying data stores can be defined for each query by the extended relational algebra. Another possibility to achieve semantic completeness is provided by systems such as \emph{CloudMdsQL} and \emph{BigDAWG} that allow using multiple query languages and their semantics in combination.

Overall, the characteristics of the different system architectures are similar. Almost all systems prefer a CA-approach. PolyBase is, caused by the fail-safe structure of the underlying HDFS, the only system that favours availability and partition tolerance over consistency. Solely, Polypheny-DB is the only system that enables the user to choose between consistency and partition tolerance.
As a form of data integration, most systems follow a LAV style or at least a hybrid variant. Only FORWARD and PolyBase are using the GAV approach. In FORWARD the extensibility of new data stores is handled by dedicated interfaces and already approved for AsterixDB. Only Polypheny-DB and \emph{PolyBase} do not take extensibility into account. All other systems propose individual solutions, but are as a result more complex.

The adaptivity dimension is considered by none of the existing systems. Only aspects such as providing concurrency and location transparency are shared among all systems. The former is caused by the systems' read-only nature. Logical independence is provided by each system except CloudMdsQL. Polypheny-DB, QoX, PolyBase and BigDAWG are able to migrate data during query execution at least temporarily. Only \emph{Polypheny-DB} is able to change the used data stores and distribution model (replication/partitioning), intends to orchestrate hardware and (geo-wide) data placements if changes are required.

In summary, we observe that a large amount of requirements is not suitably realized by the presented set of polyglot data stores. The system which seems to achieve the most of them is currently Polypheny-DB, but no coherent prototype or implementation is available yet. According to the authors, some of the necessary components have already been realized as standalone applications or as parts of other systems.

\subsection{Transparency}
Given by the underlying design principles of Poly- and Multistores they are complex software structures composed of multiple heterogeneous data stores in order to utilizes the different advantages of the underlying stores. As we established in Section~\ref{sec:Concluded Requirements}, it is beneficial to hide the adaption and data distribution processes so that the user can focus on working with the data itself. Therefore, the focus of this subsection is on Transparency\footnote{Clarification: A process is considered as transparent if the inner working of the process are not inherently visible from the outside.}, especially w.r.t. location transparency, transformation/migration transparency and concurrency transparency. These three key areas were chosen because of their importance for the seamless distribution and processing of data over multiple integrated data stores.

Based on the evaluation of the different Poly- and Multistore systems it became very clear that all systems hide the inner workings of their polyglot systems. No system communicate any information of the data that is linked to the location or migration of the data. For any user it is not directly distinguishable whether the processing is done on a single store system or a Poly-/Multistore. Furthermore, the systems don't show how the data is processed, how the data retrievals or migrations are handled or how they may be done concurrently.

The only notable exemption is BigDAWG. This system only provides some level of transformation/migration transparency if the query is focused on a single island. In any case where multiple islands are part of a BigDAWG query a casting mechanism has to be formulated.

\subsection{Query Interface}
One of the most defining features of Poly- and Multistores are their query interfaces. They not only specify the type of the system but also characterize the properties that this system is able to offer to its users.

The first contact the user has with the system is usually with the offered query languages. It decisively determines which data models will be used externally and how powerful the descriptive tools are the users may utilize for working with the data. 
An analysis of the implemented query languages of all evaluated systems shows two main trends: The usage of a SQL derivative/dialect and the implementation of an original query language. Systems which use their own language are mainly CloudMdsQL with its query language of the same name, Myria with MyriaL and RHEEM with RheemLatin. Furthermore, MyriaL finds usage in the Polystore BigDAWG, too, as the query language of one of BigDAWG's islands.
Nonetheless, many of these system offer further ways for the user to interact with the database. For example, all three integrate Python interfaces. In addition to this, Myria implements Datalog. Likewise, RHEEM extends its offering by a multitude of systems and APIs, like RheemStudio, Scala, Java and REST.

The datastores incorporating SQL-based languages cover a wide variety of dialects. The Google product BigIntegrator uses both SQL as well as the in-house developed Google Query Language, which has a SQL-like syntax for operating on their Google Cloud Plattform. Microsofts PolyBase employs Transact-SQL (T-SQL), developed by SAP and Microsoft. Polypheny-DB plans for at least one SQL dialect for its query interface and proposes PolySQL, which was developed as part of the ICARUS multistore by the same research group \cite{vogt2017icarus}. In addition to SQL, Polypheny-DB allows interaction via simple CRUD operations, an openCypher implementation and vitrivr, a multimedia retrieval engine\cite{rosetta2016vitrivr}. A quite unique approach is the query language used in FORWARD, SQL++, which is an extension of SQL to include NoSQL data models and JSON. Additionally, the database system AsterixDB uses SQL++, too, to enable users to work with its own data model \emph{ADM} (Asterix Data Model), which is based on a superset of JSON \cite{alsubaiee2014asterixdb}.

ESTOCADA is a notable exception as it does not offer a common query language at all. Instead, it offers a pivot language based on relational algebra which encapsulates the query languages of its underlying data stores. ESTOCADA offers a query languages in a similar fashion by incorporating the native query languages. Additionally, its language portfolio is extended by the Python scripting language.\\

An observation that is very apparent after a while comparing and evaluating poly- and multistores is the lack or omission of a Definition or Manipulation Language in the sources.
In the case of a \emph{Definition Language}, only Polypheny-DB, PolyBase and FORWARD are explicitly offering any kind of possibility to define schemata. All three stores use their query interface languages for this (e.g. PolyBase $\rightarrow$ T-SQL, FORWARD $\rightarrow$ SQL++). 

For \emph{Manipulation Languages}, these three stores are joined by CloudMdsQL. The data manipulation is done by using the native languages of the underlying stores that are offered by the poly- and multistore through their query interfaces or - in the case of FORWARD - by SQL++. Additionally, Polypheny-DB's design should allow to make the updated in-place or versioned.\\

One important factor when dealing with a mediation layer like a poly-/multistore is the capability of the store to preserve the Functional Completeness of Incorporated data stores through their query interfaces.
Three of the ten systems we evaluated provided the functional completeness by the native query languages of underlying stores, namely CloudMdsQL, ESTOCADA and Polypheny-DB. BigDAWG uses a similar tactic by offering so-called degenerate islands, which are only which encapsulate only a single store and its query language.
BigIntegrator, Myria, FORWARD and RHEEM solve this issue by allowing user-defined function, thus offering the possibility to manually extend the multi-/polystore by the desired functional capabilities. It should be noted, that FORWARD's query language SQL++ does not enable \emph{direct} access of the native store functionality. In PolyBase a subset of T-SQL operations (e.g. provided by the Hadoop DB Bridge) offers a portion of the functional completeness of the stores connected to the SQL server. MusQLE lacks any mechanism to preserve its stores' functionalities.\\

The \emph{Input Models} the stores expect are very dependant their implemented query languages. Accordingly, it not very surprising that the prominence of relational languages also translate to a high number of stores (BigIntegrator, Myria, PolyBase, MusQLE) that work with a relational model. In addition, FORWARD uses a relational model which is extended to provide JSON-support. CloudMsdQL and Polypheny-DB are flexible on their input model as it depends on the used query language. Similarly, the expected input model in BigDAWG is dependent on the used Island. ESTOCADA used a fixed defined list of four data models, which include a nested relational model, the key-value model, graphs and documents. Finally, RHEEM uses a unique approach in the form of data quantas, which is an abstraction of tuples, graph edges and document contents.

\subsection{Adaptivity}

In order to enable the polyglot system to react to changing requirements impose by the user directly or by changing workloads it has to adapt itself. In Section~\ref{subsec:Functional Features} on page~\pageref{inpara:Adaption level} different levels of adaptivity have already been presented. Each level corresponds to a different kind of system adaption with a severity regarding costs and extent of necessary system changes. In this section we want to focus on three topics that we consider important for adaption in a Poly- or Multistore: Logical data independence, cross-model persistence and used adaption approaches.

If we look at DBMS design, one key feature of most database systems is data independence. This feature ensures that changes in lower layers won't affect the higher ones. If we deal with adaptivity the question of how adaptive a system can be quickly correlates with the questions on how data independent the Poly-/Multistore layer is to the underlying data stores.
Most of our evaluated systems imposed logical data independence between the polyglot system layer and the incorporated data stores but there were two notable exceptions: \emph{CloudMdsQL} and \emph{BigDAWG}.
While \emph{BigDAWG} technically ensures data independence, it only does this on its island level (so within an island). As soon as the island context is left, no data independence is warranted. \emph{CloudMdsQL} doesn't implement any form of logical data independence what so ever between its mediation layer and its data stores.

In order to store data in these Poly- and Multistores in a polyglot way, these system employ generally a static cross-model persistence approach. This means that the location of data and the sued data model is mostly predetermined/defined and then appropriately distributed on run-time. This distribution can be given in the query statement itself like in CloudMdsQL where the used native query language implicitly determines the used store. Alternatively it can be given by a explicit schema mapping, a method i.e. Polybase employs. However, these static methods do not distribute whole data entities by e.g. fragment the data horizontally or vertically. Nevertheless, the horizontal and vertical distribution of data is by no means something that does not find application. In our evaluation both Polypheny-DB and MuSQLE are system which incorporate these strategies in their cross-model persistence approach.

During our evaluation one aspect of currently available Poly- or Multistore systems became quite apparent: Nearly no system incorporates any real form of adaptivity. Given the sheer amount of challenges the development a Poly-/Multistore has on its own, the addition of a (sophisticated) adaption system on top may not be the focus for the involved developers. Nevertheless, research projects such as \emph{Polypheny-DB} show that systems potentially will develop in this direction in the future. This system is the only system which is incorporating adaption into its system design. Triggered by estimations of current and future workloads \emph{Polypheny-DB} should adapt by adding or removing data stores to its running system as well as migrate and replicate data to adjust data locality and distribution. At this point, however, it must unfortunately still be pointed out that \emph{Polypheny-DB} has so far only described this function in a vision paper but has not implemented it.
A system which implements a practical and functional form of adaptivity in our evaluation is \emph{RHEEM}. It is the only other system which describes adaptivity as part of its systems functionality. \emph{RHEEMs} adaptivity shows itself in the form of a cost model learner, which adapts the cost function for the query processing by tracing the resource consumption when processing data objects (cf.~Section~\ref{subsec:RHEEM}).

\subsection{Additional Features}
Additional features describe extra functionalities to the actual data processing that are supported by the system \emph{independently}. These functionalities are relevant for the security operationalization within the e-Commerce use cases in Section~\ref{sec:UseCaseECommerce} to restrict access for the users. Our system comparison shows that systems such as BigDAWG, CloudMdsQL, ESTOCADA, Myria do not provide native support for data security or privacy to store or control access to data securely. This access must be specified externally by the deployment or network systems. Myria's documentation describes how to deploy in a cloud environment for Amazon Web Service to control a virtual firewall for inbound and outbound traffic using a \emph{security group}. However, internal access controls are missing. Polypheny-DB focuses on implementing the General Data Protection Regulation for geographic distribution in its paper but does not describe any specific action. PolyBase connector for Microsoft SQL Server uses the security features of the intercepted SQL server. 
In addition to production relevant security, push-based queries are relevant for stream-based processing. None compared system supports this feature in which queries cannot be registered with the system and returned to the caller as the response when data states changes. This lack in query registration has its reason for not updating existing datasets on the level of the mediating system like BigIntegrator or MuSQLE, where the focus lies in integrating various sources in a read-only manner.
Any comparative systems do not offer the ability to define a schema. Myria allows the sending and reading of \emph{small} data sets, for which an input schema with \emph{attribute names} and \emph{type} can be specified. Further annotations are not possible or used by the mediating system.

\subsection{Query Processing}

For the most part, query processing in the systems is viewed very differently. Starting from the \ref{tab:QueryProcessing} table, we consider five distinguishing criteria in processing. Semantic closure describes how the system handles the diverging semantics of the data models and underlying query languages, such as the behaviour with NULL values or missing attributes. 
While BigIntegrator and RHEEM have their semantics for predefined operations fixed for RHEEM's RHEMLatin language, among others, FORWARD allows changing the operational semantics. Employing possible annotations to the SQL++ queries, the responsibility is shifted to the user, who chooses the interpretation of operators or above mentioned NULL values. This adaptability on the user level also supports MuSQLE. While a standards-compliant SQL interface defines many operations upfront, it extends them with store-specific capabilities in the form of available functions. CloudMdsQL and BigDAWG refine the use of native query languages by embedding them in a language for parts of the grammar this specific language can contain. This non-agnostic system approach means that individual semantics must be dealt with within the system interaction since the BigDAWG and CloudMdsQL do not translate anything more. The behaviour of a query is dependent on the store executing it. The input and output are defined restrictively using migration from a language or semantic environment (for example, an Island within BigDAWG). CloudMdsQL proposes a fixed input and output typing via relations, whereas BigDAWG needs an additional migration step by the user. Complete semantic transparency is achieved in PolyBase. Any operations and queries are defined as Transact-SQL query, whose language semantics is already specified and executed by SQL server mediator in case of differences if needed. PolyBase avoids structural, semantic mismatches such as NULL value handling or missing attributes because the mapping of the external sources must be defined before. The user maps external sources to relations, resolves incompatibilities of types and structures (e.g. flatten hierarchical sources), and deals with missing values in the previous step.

When comparing the query processing feature of the considered multi/polystore system, our analysis shows that multi/polystores have different planning and execution approaches and leave some planning to the user. They all have in common that the system decompose incoming queries into sub-subqueries, each assigned to a different platform. The partial results are merged into a uniform format using a preferred join approach. BigIntegrator, ESTOCADA, Myria, PolyBase, FORWARD, MuSQLE, and RHEEM decompose a global query to those systems where data is located at the time of the query. Knowledge of current data locality, however, is not managed by everyone. CloudMdsQL shifts the knowledge of which platform contains the desired data to the user. Primary, the user already described the decomposition by the CloudMdsQL query language, and the system merges each subset, using the bind-join into a relational format. Compared to this, BigDAWG delegates the query to a planning component. The Planner optimizes for semantically equivalent operations. Previous benchmarks from a training phase allow the Planner to choose the right engine within the island language capsule (e.g. SQL) and limit migration only on these language and API levels.
PolyBase also looks for a good execution plan for sub-branches of a query. The goal is to leave as large a branch of the query tree as possible in the databases, pushing down the operations in the tree into systems where the data resides. However, compared to BigDAWG, PolyBase does not have the approach of moving data between platforms. Data is loaded as needed in the PolyBase Mediator when a particular operation is impossible (e.g. a JOIN operation within a MongoDB). For planning in the SQL server where PolyBase settled, all data is managed in a relational format such as MuSQLE and ESTOCADA, requiring a previous mapping configuration. Instead of only considering migrations within a subset of engines sharing the same API or language as BigDAWG, RHEEM provides additional missing migration between platforms within the complete query. In combination with the unit cost of each operation, RHEEM determines necessary migration between platforms and looks up cost-efficient transformation paths within an available conversion graph and tries to select those platforms based on learned workload from previous executions. This learning-based and dependency resolution approach makes data placement dynamic on multiple platforms. RHEEM uses an extra conversion graph to transform data objects from one platform to another, even if a conversion is only available by shifting them to an intermediate platform first. Completely without model transformations, ESTOCADA, Myria and MuSQLE achieve processing. MuSQLE builds on the relational model and makes use of relational optimization. MuSQLE enumerates and selects the cost-efficient relational query. A graph search on the relational plan tries to find a suitable join plan, with connected subgraphs, selecting the platforms where these relational subqueries shall be applied. Each platform selection is determined by comparing potential data locality with the actual data placement on a platform. MuSQLE selects those platforms for a subquery by utilizing static cost functions or querying the cost estimation of the underlying platform. Since the context remains in the relational model, the effort in the transformation is low but mainly characterized by the transfer costs. Comparing MuSQLE with FORWARD and ESTOCADA, data is moved between platforms, but all of them keeps data within a relational context. FORWARD and ESTOCADA extends the relational model, planning and executing queries on views. The mediator maintains the state of the views on each platform or incrementally updates their materialization. Migrations within query processing do not take place.

\subsection{Migration}
Migration moves and transforms data objects from one or more sources to more targets, preserving the information content. In the context of Multi-/Polystores, data movement is supported in CloudMdsQL, Myria, Polypheny, PolyBase, BigDAWG, MuSQLE and RHEEM. FORWARD and BigIntegrator are the only systems without any described migration task. They focus on closing the gap of the semantics differences between models by empowering query language or restricting expressive. FORWARD extends the syntax of the query language, considering the semi-structured facets, whereas BigIntegrator uses GQL and reduces queries only to use filters.
In contrast, other systems concern the migration of datasets directly within the system and planning phase or have to be addressed by the user. These levels differ significantly in the transparency of migrations in the systems, conducted in multiple ways. BigDAWG forces the user to describe migrations within a query to manually resolve model dependency and differences. The available CAST operator can embed the original question, defining the translations from a given schema to another selected platforms/islands (PostgreSQL schema, SciDB schema, and text schema).
In comparison, RHEEM allows an automatic transformation across multiple models. They use the conversion graph, which contains vertices of data models and edges with necessary transformations. The graph allows resolving single source multiple target paths to convert a source into required targets of an operation and their applied platforms, enabling to switch from one platform to another. PolyBase and MuSQLE pursue this in on-demand strategy and query planning. While RHEEM adds the additional migration steps into the query plan afterwards, MuSQLE does not calculate any migration costs but the execution within the plan using a given cost model. MuSQLE moves data, contrary to transfer object for an operation completion such as a JOIN, coming from preceding steps (e.g. a selection). Since MuSQLE mediates data on multiple relational systems, data model transformations are unnecessary compared to RHEEM. PolyBase, on the other hand, transfers data only when adding or as needed when the push-down of an operation is not possible.

\section{Open Challenges of Multi- and Polystores}\label{sec:Open Challenges}
As our comparison in Section~\ref{sec:Comparison} shows, there is a whole bunch of remaining challenges.
A rich set of functional features is required. Thus, full read and write support is indispensable. However, query expressiveness (e.g. complex predicates, full joins, or aggregations) may be subject of a trade-off with the level of flexibility/adaptivity, the polyglot data store supports in terms of dynamic changes of the underlying system's topology. Users must be able to specify a (database) schema for their application requiring an appropriate DDL. Furthermore, it should be possible to use the system in the traditional pull-based way but also in a push-based style with sufficient support for data privacy and security.

Concluding, we see that the main challenges will be the following:
 \begin{enumerate}[label=(\roman*)]
     \item How can initial user/application requirements be mapped to capabilities of data stores and be derived as an appropriate initial topology?
     \item How to maintain a mapping with a changeable data store topology and schema?
     \item When should an adaption process be triggered and to which extent (e.g. Temporary or persistent data migration, Adding or removing data store to/from the topology)?
     \item How can the individual store capabilities be incorporated into the query processing while keeping their semantic completeness?
     \item How can we monitor a polyglot system to enable autonomous decisions regarding adaption?
 \end{enumerate}

We consider it highly interesting to find out, whether or not the performance gains, which can be reached by topology changes, can be significantly higher than the overhead which must be spent in order to provide full adaptivity. Thus, we aim at the conceptualization, realization and evaluation of a polyglot persistence mediator supporting full adaptivity.

\section{Related Work}\label{sec:Related Work}

The emergence of NoSQL data stores since the late 2000s led to a multitude of database systems with their respective sweet spots and limitations. This landscape offers a wide range of opportunities to gain performance and functionality by choosing the right store for their needs. However, the amount of stores makes it difficult to keep an overview of all of them. For this reason, efforts were made in categorizing them, specifying their properties and providing a decision guide \cite{gessert2017nosqltoolbox}. Unfortunately, this paper only focuses on single data store environments and do not cover the recent developments in the context of integrated polyglot settings.

The general idea of the previously valid \emph{One size fits all} paradigm has been discussed by \citet{stonebraker2005onesizeI} and extended with future data management solutions and research directions in \cite{stonebraker2007onesizeII} (e.g. using abstract data types or data federation).

Efforts to categorize the landscape of the emerging polyglot data stores have only been done by a few researchers so far. \citet{tan2017enabling} focus on examining properties such as heterogeneity of query interfaces or data stores, flexibility regarding schema or extensibility, the autonomy of the underlying data stores, transparency and optimality. The work of \citet{bondiombouy2016query} targets query processing techniques and query interface capabilities of multistore systems. Properties such as adaptivity, system changes during runtime or topology adjustments have not been covered before. 

\section{Conclusion}\label{sec:Conclusion}
In this paper, we addressed the issue of polyglot data stores for two representative use cases. Based on these, we outlined the need of systems capable of tackling a multitude of (even conflicting) requirements, which can not be provided by using only a single store technology in the backend. Therefore, we presented a selected set of Poly- and Multistores designed to incorporate a collection of heterogeneous data stores and evaluated their design/capabilities regarding our outlined requirements (Tables~\ref{tab:Architecture}\nobreakdash-\ref{tab:FuncFeatures}).

We concluded, that due to their static underlying data store landscape current Poly- and Multistores show a lack of adaptivity in their systems. In addition, most of the systems are not able to exploit the unique capabilities of their incorporated data stores. In our opinion, adaptivity is the key feature to guarantee future-proof, long-term usability of polyglot data stores in a world of ever changing application and data requirements. This results in a set of challenging research question regarding adaptivity as outline in Section~\ref{sec:Open Challenges}.\\

Based on our expertise in the fields NoSQL database technology \cite{gessert2017nosqltoolbox} and scalable data management \cite{gessert2017quaestor}, we aim at an automated polyglot persistence mediator which propose a declarative approach allowing the application to initially specify an annotated data schema \cite{schaarschmidt2015automatedpolyglot}. It should be possible to annotate schema elements by continuous non-functional (e.g. a certainly acceptable write latency), binary non-functional (e.g. atomic updates), or binary functional (e.g. consistency level read-your-writes required) requirements to express SLAs. Depending on the functional and non-functional capabilities of (NoSQL) systems, a smart algorithm is supposed to rank available systems and derive the best possible topology including a routing model which defines the mappings from schema elements to databases.

\bibliographystyle{ACM-Reference-Format}
\bibliography{bibliography/own_literature, bibliography/main_literature}

\end{document}

%% file: pictures/system_types.tex
\newcommand{\multimodel}[1]{\includegraphics[page=1, trim=65 92 390 150, clip, #1]{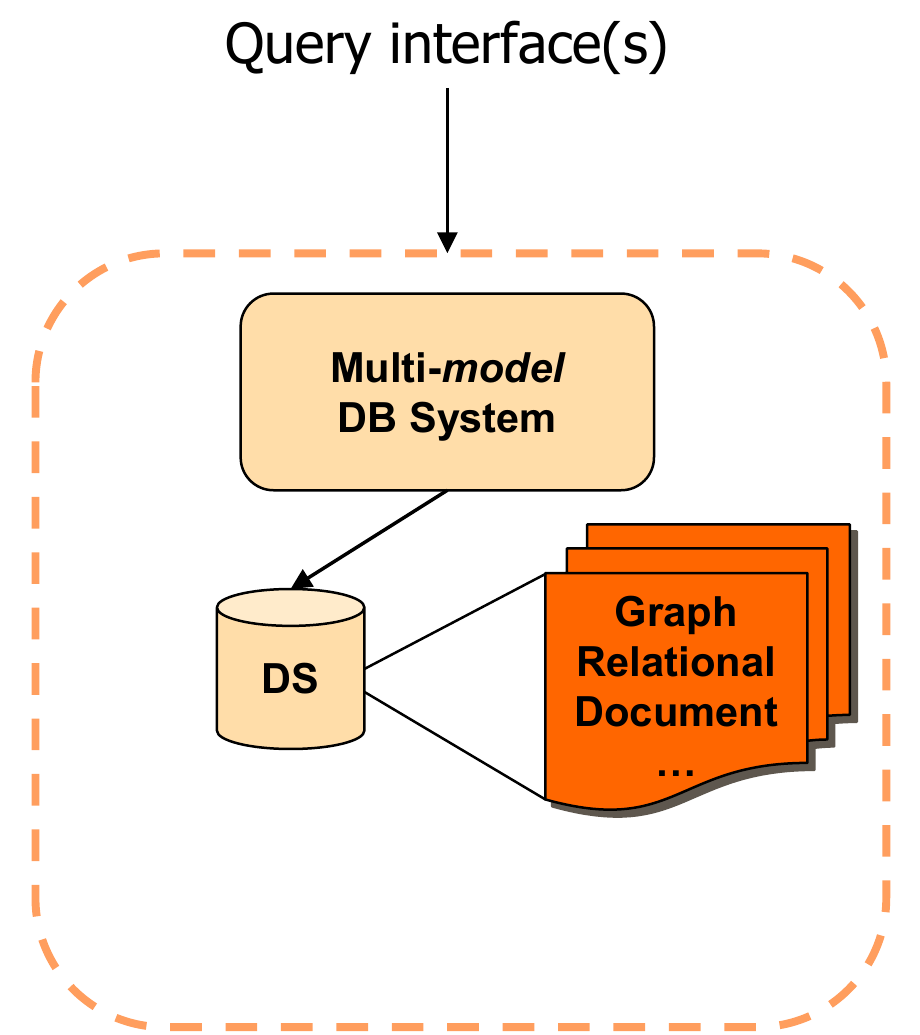}}

\newcommand{\multimodal}[1]{\includegraphics[page=1, trim=395 92 65 150, clip, #1]{./pictures/system_types_v2.pdf}}

\newcommand{\federated}[1]{\includegraphics[page=2, trim=65 92 390 150, clip, #1]{./pictures/system_types_v2.pdf}}

\newcommand{\polylingual}[1]{\includegraphics[page=2, trim=395 92 65 150, clip, #1]{./pictures/system_types_v2.pdf}}

\newcommand{\multistore}[1]{\includegraphics[page=3, trim=65 92 390 150, clip, #1]{./pictures/system_types_v2.pdf}}

\newcommand{\polystore}[1]{\includegraphics[page=3, trim=395 92 65 150, clip, #1]{./pictures/system_types_v2.pdf}}

%% file: pictures/figure_system_types.tex
\newcommand{\subCaptionTextSize}{\small}

\newcommand{\columnWidthFactor}{0.25\textwidth}

\begin{figure}[t]%
    \begin{minipage}[t]{\columnWidthFactor}
        \subfloat[\subCaptionTextSize Multi-Model]{
            \multimodel{width=\columnwidth}
            \label{subfig:multimodeeel}
        }
        
        \subfloat[\subCaptionTextSize Multi-Modal]{
            \multimodal{width=\columnwidth}
            \label{subfig:multimodaaal}
        }
    \end{minipage}%
    \begin{minipage}[t]{\columnWidthFactor}
        \subfloat[\subCaptionTextSize Federated]{
            \federated{width=\columnwidth}
            \label{subfig:Federated}
        }
        
        \subfloat[\subCaptionTextSize Polylingual]{
            \polylingual{width=\columnwidth}
            \label{subfig:Polylingual}
        }
    \end{minipage}%
    \begin{minipage}[t]{\columnWidthFactor}
        \subfloat[\subCaptionTextSize MultiStore]{
            \multistore{width=\linewidth}
            \label{subfig:MultiStore}
        }
        
        \subfloat[\subCaptionTextSize PolyStore]{
            \polystore{width=\linewidth}
            \label{subfig:PolyStore}
        }
    \end{minipage}%
    \caption{System Types (Based on: Tan et al. \cite{tan2017enabling})}
    \label{fig:SystemTypes}
\end{figure}

%% file: tables/table_commands.tex
\definecolor{jaGreen}{rgb}{0.227, 0.588, 0.016}
\definecolor{jeinYellow}{rgb}{1, 0.55, 0}
\definecolor{unknownBlue}{rgb}{0.1721, 0.2219, 0.6060}

\newcommand{\customTableWidth}{1.8\textwidth}
\newcommand{\customResizebox}{1.0\textwidth}

\setlength{\rotFPtop}{0pt plus 1fil}
\setlength{\rotFPbot}{0pt plus 1fil}

\newcommand{\tabNein}[1][]{\textbf{\textcolor{red}{\texttimes}}#1}
\newcommand{\tabJa}[1][]{\textbf{\textcolor{jaGreen}{\checkmark}}#1}
\newcommand{\tabJein}[1][]{\textbf{\textcolor{jeinYellow}{\textcircled{\scriptsize\checkmark}}}#1}
\newcommand{\tabUnknown}[1][]{\textbf{\textcolor{unknownBlue}{\textcircled{\scriptsize ?}}}#1}

\newcommand{\footref}[1]{\textsuperscript{\ref{#1}}}

\renewcommand{\tabularxcolumn}[1]{m{#1}} 
\newcolumntype{C}{>{\centering\arraybackslash}X} 
\newcommand{\tableFontSize}{\normalsize} 
\newcommand{\tableFootnoteFontSize}{\tiny}
\newcommand{\tfnfs}{\tableFootnoteFontSize}

%% file: tables/table_architecture.tex
\begin{sidewaystable}[htp]
\centering
\tableFontSize
\caption{Categorization and comparison of the systems from section~\ref{sec:Analyzed Systems}\\(\tabJa~-~existent, \tabNein~-~nonexistent, \tabJein~-~partially fulfilled, \tabUnknown~-~unknown)}
\label{tab:Architecture}
\resizebox{\customResizebox}{!}{%

\begin{tabularx}{\customTableWidth}{C|C||C|C|C|C|C|C|C|C|C|C}
\hline
 \multicolumn{2}{c||}{} & \textbf{CloudMdsQL} & \textbf{BigIntegrator} & \textbf{ESTOCADA} & \textbf{Myria} & \textbf{PolyphenyDB} & \textbf{Polybase} & \textbf{BigDAWG} &  \textbf{FORWARD} & \textbf{MuSQLE} & \textbf{RHEEM}\\
 
 \hline
 \hline
 
 \multicolumn{2}{c||}{\textbf{System Type}} & Multistore & Multistore & Multistore & Multistore & Polystore & Multistore (HDFS-Bridge) & Polystore & Multistore & Multistore & Multistore \\ \hline
 
 \multicolumn{2}{c||}{\textbf{System Availability}} & research system & research system & research system & research system & vision & commercial & research system & research system & research system & research system \\ \hline
 
 \multicolumn{2}{c||}{\textbf{System Evaluation}} & \cite{kolev2016benchmarking} & \tabNein & \cite{alotaibi2019towards} & \cite{wang2017myria, halperin2014demonstration} & \tabNein & \cite{dewitt2013split} & \cite{karimov2018polybench, yu2017database} & \tabNein & \cite{giannakouris2016musqle} & \cite{agrawal2016rheem}\\
 
 \hline
 \hline
 
 \multirow{5}{*}{\textbf{Architecture}} & \textbf{Component connection} & hybrid-coupled & loosely-coupled & tightly-coupled & loosely-coupled & \tabUnknown & loosely-coupled & hybrid-coupled & loosely-coupled & tightly-coupled & tightly-coupled \\ \cline{2-12}
 
 & \textbf{Mapping} & Hybrid & LAV & LAV & LAV & \tabUnknown & GAV & Hybrid & GAV & GAV & LAV \\ \cline{2-12}
 
 & \textbf{Extensibility} & \tabJa (+) & \tabJa (++) & \tabJa (++) & \tabJa (+) & \tabUnknown & \tabUnknown & \tabJa (+) & \tabNein & \tabJa (+) & \tabJa (+) \\ \cline{2-12}
 
 & \textbf{CAP Orientation} & CA & CA & CA & CA & CP & AP & CA & CA & CA & CA \\ \cline{2-12}
 
 & \textbf{Currently supported data stores} & PostgreSQL (JDBC), MongoDB, Apache Spark, Derby, Pyhton\footnote{No data store} & BigTable, standard-SQL supporting DBMS & relational systems, Spark, AsterixDB, Redis, Solr, Saxon & SciDB, HDFS, MyriaX, SPARQL supporting stores & MariaDB, PostgreSQL, VoltDB, MonetDB \footnote{Set of DBMS currently supported by Icarus implementation} & Microsoft SQL Server, Oracle, Teradata, MongoDB, Generic ODBC, Hadoop, Azure Blob Store & PostgreSQL, SciDB, Accumulo, S-Store, Myria & AsterixDB, MongoDB, CouchDB & PostgreSQL, MemSQL, SparkSQL & JavaStreams, PostgreSQL, Flink, Spark, GraphChi, Giraph \\
 
 \hline
 \hline
 
 \multirow{3}{*}{\textbf{Transparency}} & \textbf{Location} & \tabJa & \tabJa & \tabJa & \tabJa & \tabJa & \tabJa & \tabJa & \tabJa & \tabJa & \tabJa \\ \cline{2-12}
 
 & \textbf{Transformation / Migration} & \tabJa & \tabJa & \tabJa & \tabJa & \tabJa & \tabJa & \tabJein\footnote{Only for single-island queries.} & \tabJa & \tabJa & \tabJa \\ \cline{2-12}
 
 & \textbf{Concurrency} & \tabJa & \tabJa & \tabJa & \tabJa & \tabJa & \tabJa & \tabJa & \tabJa & \tabJa & \tabJa \\ \hline

\end{tabularx}






}
\end{sidewaystable}

%% file: tables/table_functional_features.tex
\begin{sidewaystable}[htp]
\centering
\tableFontSize
\caption{Categorization and comparison of the systems from section~\ref{sec:Analyzed Systems}\\(\tabJa~-~existent, \tabNein~-~nonexistent, \tabJein~-~partially fulfilled, \tabUnknown~-~unknown)}
\label{tab:FuncFeatures}
\resizebox{\customResizebox}{!}{%

\begin{tabularx}{\customTableWidth}{C|C||C|C|C|C|C|C|C|C|C|C}
\hline
\multicolumn{2}{c||}{} & \textbf{CloudMdsQL} & \textbf{BigIntegrator} & \textbf{ESTOCADA} & \textbf{Myria} & \textbf{PolyphenyDB} & \textbf{Polybase} & \textbf{BigDAWG} &  \textbf{FORWARD} & \textbf{MuSQLE} & \textbf{RHEEM}\\
 
\hline
\hline
 
\multirow{6}{*}{\textbf{Query Interface}} & \textbf{Query Language} & CloudMdsQL (using native query languages), Python & SQL, GQL & No common query language; pivot language: relational algebra & Datalog, MyriaL, Python & SQL dialect (e.g. PolySQL\cite{vogt2017icarus}), CRUD, openCypher, vitrivr & T-SQL & SQL, AFL MyriaL, SPARQL (island-dependent) & SQL++ & SQL & RheemLatin, TheemStudio, Python, Scala, Java, REST\\ \cline{2-12}

& \textbf{Definition Language} & \tabNein & \tabNein & \tabNein & \tabNein & SQL dialect \footnote{Icarus uses self-developed dialect PolySQL \cite{vogt2017icarus}}, CRUD interface, openCypher, vitrivr & T-SQL & \tabNein & SQL++ with JSON related Types and Datasets (in AsterixDB) & \tabNein & \tabNein\\ \cline{2-12}

& \textbf{Manipulation Language} & Data changes defined as native queries & \tabNein & \tabNein & \tabNein & possible in all available query languages; updates can be made in-place or versioned & T-SQL & \tabNein & SQL++ with INSERT/ DELETE/ UPSERTS of datasets (in AsterixDB) & \tabNein & \tabNein\\ \cline{2-12}

& \textbf{Functional Completeness} & Provided by native query languages & only by providing extending query execution plugins & Native queries allow native feature access & Allows user-defined functions and aggregates, constrained imperative & Provided by native query language & Subset of T-SQL operations, provided by DB bridge & Provided by degenerate islands & Allows the definition of own functions; no direct access of native store functionality & \tabNein & Fixed but extendable by user-defined functions\\ \cline{2-12}

& \textbf{Input Models} & Depends on native query language used & Relational Model & Nested relational model, key-value, graphs, document & Relational Model & Depends on native query language used in subquery block & Relational Model & Island-dependent & Hybrid Model (Relational + JSON) & Relational Model & Data quanta (abstraction of tuples, graph edges, document content)\\ 



\hline
\hline

\multirow{3}{*}{\textbf{Adaptivity}} & \textbf{Logical Data Independence} & \tabNein & \tabJa & \tabJa & \tabJa & \tabJa & \tabJa & Only within islands & \tabJa & \tabJa & \tabJa\\ \cline{2-12}

& \textbf{Cross-Model Persistence} & static & static & static & static & horizontally and vertically & static & static & static & vertically and horizontally fragmented & static\\ \cline{2-12}

& \textbf{Adaption Approach} & \tabNein & \tabNein & \tabNein & \tabNein & Estimation of current and future workloads to adapt by adding/removing datastores and migrate/replicate data & \tabNein & \tabNein & \tabNein & \tabNein & Cost model learner adapts cost function for query processing\\

\hline
\hline

\multirow{2}{*}{\textbf{Additional features}} & \textbf{Data Privacy and Security} & \tabNein & \tabNein & \tabNein & \tabNein & GDPR compliant data distribution, data safety aware replication & \tabNein & \tabNein & \tabNein & \tabNein & \tabNein\\ \cline{2-12}

& \textbf{Push-based Query Execution} & \tabNein & \tabNein & \tabNein & \tabNein & \tabNein & \tabNein & In conjunction with S-Store & \tabNein & \tabNein & \tabNein\\

\hline
& \textbf{Custom Optimization Targets} & \tabNein & \tabNein & \tabNein & \tabNein & \tabNein & \tabNein & \tabNein & \tabNein & \tabNein & \tabNein\\

 
\end{tabularx}

}
\end{sidewaystable}

%% file: tables/table_query_processing.tex
\begin{sidewaystable}[htp]
\centering
\tableFontSize
\caption{Categorization and comparison of the systems from section~\ref{sec:Analyzed Systems}\\(\tabJa~-~existent, \tabNein~-~nonexistent, \tabJein~-~partially fulfilled, \tabUnknown~-~unknown)}
\label{tab:QueryProcessing}
\resizebox{\customResizebox}{!}{%

\begin{tabularx}{\customTableWidth}{C|C||C|C|C|C|C|C|C|C|C|C}
\hline
\multicolumn{2}{c||}{} & \textbf{CloudMdsQL} & \textbf{BigIntegrator} & \textbf{ESTOCADA} & \textbf{Myria} & \textbf{PolyphenyDB} & \textbf{PolyBase} & \textbf{BigDAWG} &  \textbf{FORWARD} & \textbf{MuSQLE} & \textbf{RHEEM}\\
 
\hline
\hline

\multirow{6}{*}{\textbf{Query Processing}} & \textbf{Semantic Completeness} & Semantic of native stores transformed into provided CloudMdsQL typesystem & fixed & Distinct semantics when smeantics treated differently. Encode the semantics same as much as can & Relational semantics extend by iterative processing constructs such as do-while and while loop & \tabUnknown & PDW filters for unsupported types and handles them itself & Island-dependent (defined during development of island) & varying semantics bypassed by annotations per query & Standard SQL support extended by store-specific functions & fixed\\ \cline{2-12}


& \textbf{Cost Estimation} & cost-based & \tabUnknown & dynamic & & heuristic & cost-based & heuristic, benchmarking & \tabUnknown & cost-based & adaptive cost-based\\ \cline{2-12}

& \textbf{Optimization Approach} & heterogeneous join (bind join), operator reordering & heterogeneous join (bind join), pushing-down operations & query-rewriting minimization (respecting conditions), sub-query minimization & data-movement minimization, operator reordering, heterogeneous join (tributary-join) & \tabUnknown & (classical) query planning, pushing down operations & minimization of data movement, pushing-down operations, benchmarking for semi-equivalent query & caching-like approach based on materialized views & join operator reordering (finding subplans for all possible locations), data-movement minimization & data-transformation minimization, data-movement minimization, operator-reordering, adaptive cost model\\ \cline{2-12}

& \textbf{Query Decomposition} & data location dependency, according to native query fragment in global query, according to parallel processing & data location dependency, data store capability & Native queries translated into relational pivot language & data location dependency, according to parallel processing (splitting by shared-worker thread) & data location dependency, data store capability & data location dependency & according to capability, according to data island dependency & \tabUnknown & data location dependency & data location dependency\\ \cline{2-12}

& \textbf{Internal Query Model} & Relational & Relational & Relational (views with constraints) & Relational & Associative Arrays & Relational & Island-dependent & Hybrid model consisting of JSON and relational & Relational & Data quanta (abstraction of tuples, graph edges, document content)\\

\hline
\hline

\multirow{2}{*}{\textbf{Migration}} & \textbf{Data Shipping} & \tabJa & \tabNein & \tabJa\footnote{Fragments stored as materialized views.} & \tabJa & \tabJa & \tabJa & \tabJa & \tabNein & \tabJa & \tabJa \\ \cline{2-12}

& \textbf{Migration (On- and Pre-Demand)} & \tabNein & \tabNein & & \tabNein & \tabJa (not specified further) & On-Demand & \tabNein & \tabNein & On-Demand & On-Demand\\

\hline

\end{tabularx}






}
\end{sidewaystable}